\documentclass[preprint,aps,showpacs]{revtex4}                             %

\usepackage{epsfig}

\begin{document}
\input psfig.sty

\def\beq{\begin{equation}}
\def\eeq{\end{equation}}

\title{Physical Origin of the Quadrupole Out-of-Plane Magnetic Field
in Hall-MHD Reconnection}
\author{Dmitri A. Uzdensky}
\email{uzdensky@astro.princeton.edu}
\author{Russell M. Kulsrud}
\email{rmk@pppl.gov}
\affiliation{Princeton University Observatory 
and Princeton Plasma Physics Laboratory 
-- Center for Magnetic Self-Organization (CMSO), Princeton, NJ 08544}

\date{May 11, 2006}

\begin{abstract}
A quadrupole pattern of the out-of-plane component of the magnetic field
inside a reconnection region is seen as an important signature of the
Hall-magnetohydrodynamic (Hall-MHD) regime of reconnection. It has been 
first observed in numerical simulations and just recently confirmed in 
the MRX (Magnetic Reconnection Experiment) [M.~Yamada, H.~Ji, S.~Hsu, 
T.~Carter, R.~Kulsrud, N.~Bertz, F.~Jobes, Y.~Ono, and F.~Perkins, 
Phys. Plasmas 4, 1936 (1997)] and also seen in spacecraft observations 
of Earth's magnetosphere. In this study, the physical origin of the 
quadrupole field is analyzed and traced to a current of electrons that 
flows along the lines in and out of the inner reconnection region to 
maintain charge neutrality. The role of the quadrupole magnetic field 
in the overall dynamics of the reconnection process is discussed. 
In addition, the bipolar poloidal electric field is estimated and 
its effect on ion motions is emphasized. 
\end{abstract}


\maketitle

\newpage


\section{Introduction}
\label{sec-intro}

Ever since it was established that the classical Sweet--Parker reconnection 
model~\cite{Sweet-1958,Parker-1957,Parker-1963} with Spitzer resistivity is 
too slow and that the Petschek~\cite{Petschek-1964} fast-reconnection 
mechanism cannot be realized in resistive MHD with uniform resistivity~\cite
{Biskamp-1986,Scholer-1989,Uzdensky-2000,Kulsrud-2001,Malyshkin-2005}), 
theoretical studies of fast reconnection have proceeded as a competition 
between two schools of thought. The first one invokes the idea of enabling 
the Petschek mechanism by a strongly localized anomalous resistivity due to 
plasma micro-instabilities triggered when a certain current threshold is 
exceeded~\cite{Ugai-1977,Biskamp-2001,Kulsrud-2001,Malyshkin-2005}.
The second, commonly referred to as the Hall reconnection mechanism, 
relies on the two-fluid effects that become important when the reconnection 
layer becomes so thin that ion and electron motions decouple from each 
other~\cite{Sonnerup-1979,Mandt-1994,Biskamp-1997,Shay-1998a,Shay-1998b}. 
In reality, as the layer gets thinner, the Hall term becomes more and more 
important, but it is possible that the condition for anomalous resistivity 
is reached first. If this happens, then the enhanced collision rate due to 
fluctuations allows electrons to flow across the field lines instead of 
having to flow rapidly along them to preserve charge neutrality (see below). 
This will weaken or remove the Hall effect on reconnection. On the other 
hand, it may be that a stable two-fluid flow pattern is established first, 
before the instabilities that lead to anomalous resistivity are triggered. 
Then reconnection would proceed by the Hall mechanism.

In the present study we focus on the Hall-magnetohydrodynamic (Hall-MHD) 
regime of reconnection. More specifically, our main objective is to 
understand the physical origin of the quadrupole pattern of the out-of-plane 
magnetic field that arises inside the reconnection region in this regime. 
We shall call it ``the quadrupole field'' for short. The quadrupole field 
is widely accepted as one of the most important signatures of the two-fluid
effects in the reconnection process. The reason for this
is the following. Consider the simplest two-dimensional (2D)
reconnection configuration (displayed in Fig.~\ref{fig-idea}a), 
with $z$ marking the ignorable direction, and~$x$ and~$y$ forming 
the so-called reconnection plane. Let us assume that there is no 
guide field, that is, assume that the magnetic field lines above 
and below the reconnection layer lie exactly in the reconnection 
plane (``null-helicity'' reconnection). Then, in simple
resistive magnetohydrodynamics (MHD) there is no mechanism
that would produce an out-of-plane ($z$) component of the 
magnetic field anywhere in the reconnection region; this is
basically a consequence of the symmetries inherent to the
resistive MHD equations. However, when electron and ion 
flows decouple from each other, that is when two-fluid 
effects become important, those symmetries are no longer 
present, since charge carriers of different sign now move 
differently. As a result, an out-of-plane component of
the magnetic field may develop somewhere inside the layer.
Thus, the emergence of this field is a tell-tale sign of 
the transition from resistive to two-fluid (e.g., Hall) 
regime of reconnection. For example, imagine a situation 
where one studies a reconnection process
in a lab with limited diagnostic capabilities, say, with
only magnetic probes but with no ability to measure plasma
densities, temperatures, velocities, etc. Then one is not able to
determine the ion skin depth and ion Larmor radius to compare
them with the measured reconnection layer thickness. However,
a mere detection of the $z$-component of the magnetic field
in the layer will immediately and unambiguously reveal that 
one deals with a two-fluid reconnection regime.

The presence of an out-of-plane magnetic field with a quadrupole
structure in the context of the Hall-MHD regime of collisionless 
reconnection was first suggested by Sonnerup~\cite{Sonnerup-1979}
(see also the work of Terasawa~\cite{Terasawa-1983} in the context 
of tearing instability in the Earth magnetotail). Since then, the 
quadrupole field has been observed in many numerical simulations 
of collisionless reconnection~\cite
{Mandt-1994,Biskamp-1997,Shay-1998a,Shay-1998b,Wang-2000,Arzner-2001,Pritchett-2001,Breslau-2003}. 
It has also been detected in space in-situ measurements by the {\it Polar} 
and {\it Cluster} spacecraft flying through the Earth magnetotail and the 
magnetopause~\cite{Mozer-2002,Wygant-2005,Borg-2005}. 
The quadrupole magnetic field pattern has long evaded direct experimental 
detection in laboratory plasma experiments but just recently has finally 
been confirmed~\cite{Ren-2005} in the Magnetic Reconnection Experiment 
(MRX)~\cite{Yamada-1997}. 
A similar result has recently been reported in the Swarthmore Spheromak
Experiment (SSX) in a somewhat different type of a neutral sheet~\cite
{Matthaeus-2005}. 
These experimental detections raise the need for a better theoretical 
understanding of the generation mechanism for the quadrupole field, 
which in our view is still lacking despite the great wealth of numerical 
data.

Our paper is structured as follows. In Sec.~\ref{sec-QB-generation}
we propose a basic physical explanation of how an out-of-plane magnetic 
field is generated in an electron MHD (eMHD) reconnection layer and why 
it inevitably has a quadrupole pattern. We start this section by discussing 
our physical assumptions (Sec.~\ref{subsec-assumptions}); then we present 
a simple physical description of the mechanism by which the quadrupole
field is produced (Sec.~\ref{subsec-idea}); and, finally, we illustrate 
our ideas by an analytical calculation of the toroidal (e.g., out-of-plane) 
field in the X-point configuration (Sec.~\ref{subsec-Xpoint}). 
In~Sec.~\ref{sec-general} we step back from our specific example 
of~Sec.~\ref{subsec-Xpoint} and derive some general results pertinent 
to stationary incompressible ideal eMHD in 2.5 dimensions. Thus, 
in~Sec.~\ref{subsec-Bz=Phi=V} we establish proportionality between 
three important quantities: the volume per poloidal flux computed 
along a field line, the poloidal (e.g., in-plane) electron stream 
function, and the electron contribution to the toroidal magnetic field. 
In particular, we show that, as long as ion currents are neglected in 
Ampere's law and the electrons are magnetized, the toroidal magnetic 
field is constant along electron streamlines; correspondingly, the 
toroidal electron velocity has to be constant along poloidal magnetic 
field lines. This means at least that inside the reconnection layer, 
at scales smaller than the ion inertial scale (but outside the inner 
electron dissipation region), one cannot invoke the usual explanation 
for the quadrupole field as being created by differential stretching 
of the poloidal field lines due to a non-uniform toroidal electron flow.
It then follows that the toroidal field has to be generated in the
transition region in the outskirts of the reconnection layer, where
the ion-current contribution is still important. To study this process, 
we consider the three-dimensional shape of the field lines and show 
that the toroidal separation $\Delta z$ between a given fluid element 
on a line and the tip of the line at the $x=0$ plane is related to 
the volume per flux integral (Sec.~\ref{subsec-shape-xz}). 
This enables us to calculate the toroidal electron velocity 
(Sec.~\ref{subsec-v_z}) and hence estimate when (i.e., how close 
to the separatrix) electron inertia becomes important in the 
generalized Ohm law (Sec.~\ref{subsec-inertia}). Finally, we 
argue that the toroidal velocity of the field lines should
be attributed to an ${\bf E\times B}$ drift of electrons; 
this requires the presence of a bipolar poloidal electric 
field which we compute in~Sec.~\ref{subsec-bipolar-E}.
This electric field is also an important signature of the two-fluid 
effects in reconnection; in particular, it is responsible for accelerating 
ions into the reconnection layer, resulting in an effective ion heating.
We summarize our work in~Sec.~\ref{sec-conclusions}.


\section{How is the quadrupole field generated in the reconnection region?}
\label{sec-QB-generation}


\subsection{Physical Assumptions}
\label{subsec-assumptions}

First, let us discuss the physical assumptions that we adopt in this paper.
These assumptions are aimed at making the problem tractable while still 
realistic and complex enough to provide a useful physical picture of
a reconnecting current layer in the Hall-MHD regime. While doing this, 
we pay special attention to the conditions relevant to the MRX experiment.

We will be mostly interested in the inner structure of the reconnection 
layer at scales (in the direction across the layer) smaller than the ion 
collisionless skin depth defined as
\beq
d_i \equiv {c\over{\omega_{pi}}} = c \, \sqrt{m_i\over{4\pi n_e e}} \, .
\label{eq-def-d_i}
\eeq
Provided that there is some ion heating available, and in the absence 
of a strong guiding field, the ion gyro-radius, $\rho_i$, is comparable 
to~$d_i$ in this region, and so ions can be regarded as unmagnetized. 
Thus, their motion is not strongly affected by the small-scale magnetic 
structures that characterize the inner part of the reconnection layer 
considered in this paper. On these small scales, the motion of ions is 
slow and smooth; the ion density then cannot develop structure on these 
scales. We shall therefore treat ions as providing a neutralizing 
background, which, for simplicity, we shall take to be uniform. 
Also, for the most part, we shall assume them to be motionless,
that is we shall neglect the ion contribution to the electric current. 
However, as we will show, the poloidal ion current in the outer 
region of the layer actually plays an important role in the generation 
of the quadrupole field.

In contrast to ions, electrons have very small gyro-radii and are 
well magnetized everywhere except in a small vicinity of the X-point. 
Thus, it is appropriate to use the framework of Hall MHD (or electron
MHD) in most of the region under consideration. This framework is 
characterized as a two-fluid approach where the magnetic field is 
frozen into electrons but not into ions. 
An equivalent formulation is to use a generalized Ohm law
(i.e., electron equation of motion) that includes the Hall term. 
On the other hand, since we are interested in scales that are
much larger than the size of the inner electron diffusion region,
we shall, in our analysis of the generalized Ohm law, neglect
both the electron inertia term (although we shall estimate its
contribution in~Sec.~\ref{subsec-inertia}) and the resistive term 
that arises due to normal particle-particle collisions (i.e., 
classical Spitzer resistivity) or due to wave-particle collisions 
(anomalous resistivity). At the same time, we shall include the 
electron pressure gradient term, assuming, however,
that the electron pressure tensor is isotropic. This assumption
is justified if the system is not entirely collisionless.
That is, we assume that collisions are rare enough for collisional
resistivity to be negligible, but, at the same time, frequent enough 
to restore the electron pressure isotropy throughout most of the 
reconnection region. This is in fact consistent with the conditions 
encountered in the MRX experiment, where collisions are always present
at some level~\cite{Ren-2005}.

Next, due to the charge neutrality condition (valid provided that the scales 
under consideration are still much larger than the Debye length), the 
electron density has to be equal to that of the ions. Since we assume
that latter to be uniform, we require the electron density to be also 
uniform and hence the electron flow to be incompressible.

Finally, we assume that the reconnection layer is in a quasi-steady state,
that it has a translational symmetry in one ($z$) direction, and that there 
is no guide magnetic field (that is no externally imposed toroidal magnetic 
field).

Thus, from the above considerations, the set of physical assumptions 
can be summarized as {\it ideal incompressible 2.5-D steady-state 
electron~MHD} without a guide field~\footnote
{Here, the term ``2.5-D'' refers, as usual, to a situation in 
which vectors are three-dimensional but all physical quantities 
are translationally invariant in one direction.}.


\subsection{A Simple Physical Picture of the Quadrupole Field Generation}
\label{subsec-idea}

\begin{figure} [h]
\centerline
{\psfig{file=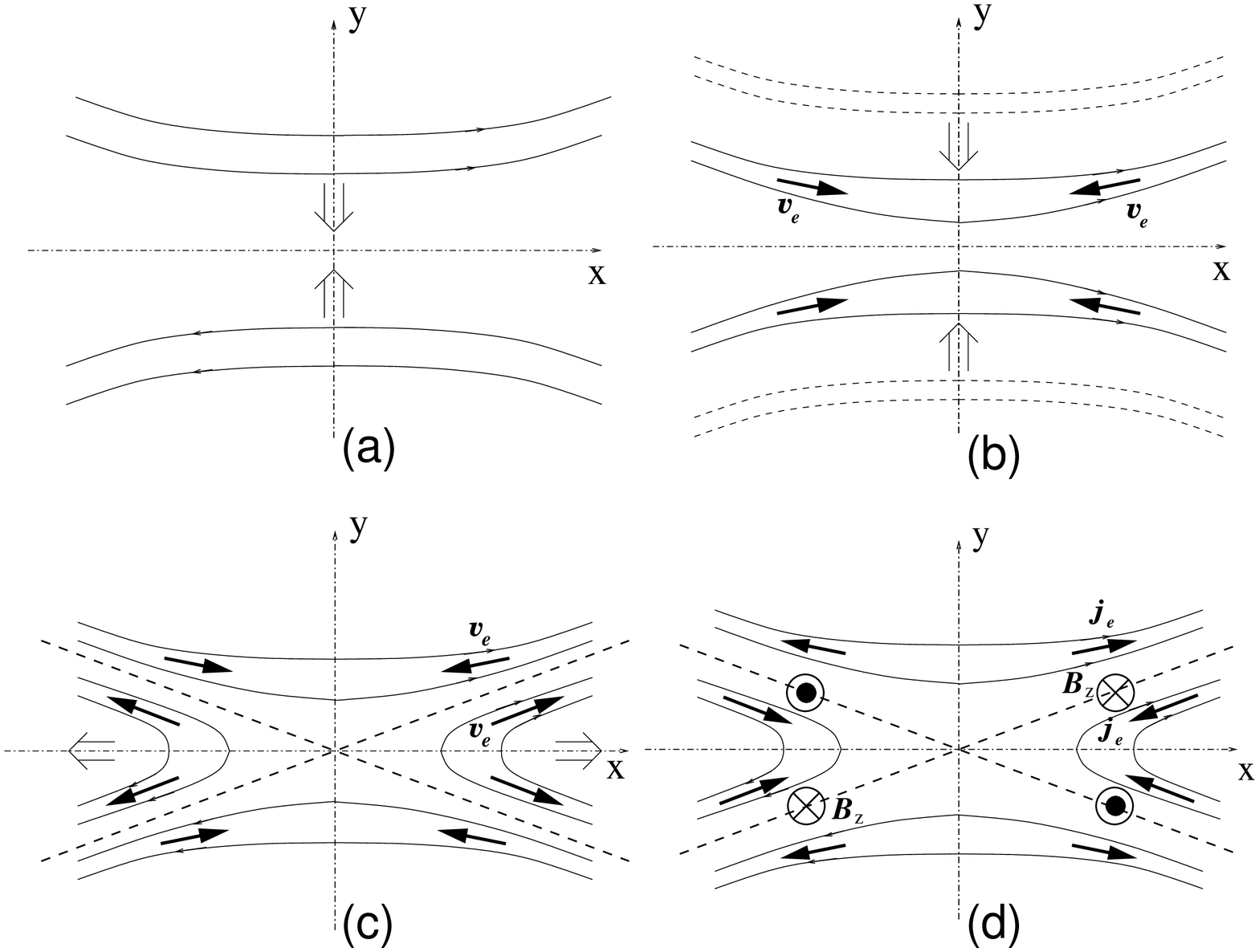,width=6in}}
\caption{The basic idea of out-of-plane field generation.}
\label{fig-idea}
\end{figure}

We first describe the basic physical picture of how the quadrupole
out-of-plane magnetic field naturally arises in electron MHD. Consider 
an incoming flux tube as it moves deeper and deeper into the (ion-scale) 
reconnection region toward the X-point (Fig.~\ref{fig-idea}a). The poloidal 
magnetic field in the central part of the tube near $x=0$ has to decrease, 
and hence the volume of this central part has to expand. Since electrons 
are tightly coupled to magnetic field lines, this expansion would lead to 
a drop in electron density. However, the ions are not magnetized and their 
density does not decrease. Therefore, since almost perfect charge neutrality 
is to be maintained, a very small poloidal electric field arises and it 
immediately pulls the electrons along the field lines inward from the outer 
parts of the flux tube into this central region. Owing to the very large 
mobility of electrons along the field (inversely proportional to~$m_e$), 
this parallel electric field is negligibly small.

As a result, we get a strong inflow of electrons along the poloidal 
magnetic field in the upstream region (Fig.~\ref{fig-idea}b). This 
inflow rapidly accelerates as the field line approaches the separatrix, 
because of the rapidly increasing rate of flux-tube expansion near the 
X-point. In the downstream region, the direction of the electron flow 
reverses: as a newly reconnected field line moves away from the X-point, 
the volume of its central part decreases and so the electrons are squeezed 
out and flow rapidly outward along the field (Fig.~\ref{fig-idea}c).
As the field line moves further away, this outflow gradually decelerates. 
The resulting overall picture of the electron flow is shown in Fig.~\ref
{fig-idea}c; once again, the main feature is the rapid inflow of electrons 
just above the separatrix followed by a rapid outflow just below the 
separatrix. 

This pattern of electron motion plays an important role
in eMHD reconnection, since there is a poloidal electric
current associated with the flow of electrons. By Ampere's law,
this current generates a quadrupole toroidal magnetic field 
concentrated along the separatrix (see Fig.~\ref{fig-idea}d). 
This is our picture for the origin of the quadrupole field.

The orientation of this field is always such that the toroidal 
field in the upper right and lower left quadrants is directed 
away from the viewer, whereas the toroidal field in the lower 
right and the upper left quadrants is directed towards the viewer. 
It is interesting to note that this orientation is universal, i.e., 
independent of the direction of the poloidal field.


\subsection{An Analytical Example: a Simple X-point Configuration}
\label{subsec-Xpoint}

To illustrate this mechanism, we present a very simple calculation 
of the toroidal field based on the simplest possible poloidal field 
configuration relevant to the reconnection problem. This configuration
is of course the X-point configuration that we now describe.

\begin{figure} [h]
\centerline
{\psfig{file=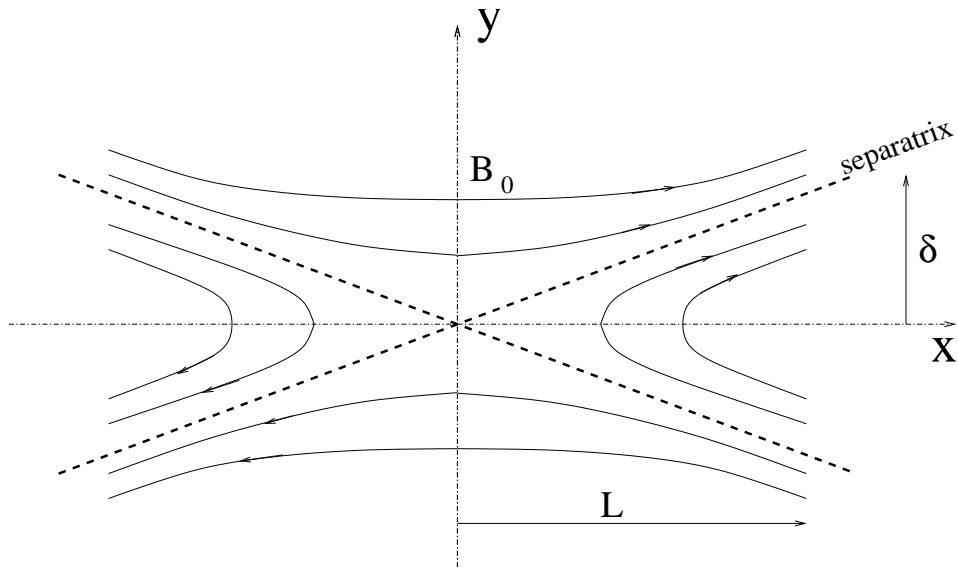,width=5in}}
\caption{Simple X-point configuration.}
\label{fig-x-point}
\end{figure}

Consider the central part of a reconnecting current layer.
Let $L$ be the half-width and $\delta\ll L$ be the half-thickness
of the layer. Let us choose a Cartesian coordinate system
$(x,y,z)$ with~$x$ being the direction along the layer, 
$y$ across the layer, and~$z$ in the ignorable direction.
We shall refer to the plane of reconnecting field (i.e., 
the $xy$ plane) as the poloidal plane and the $z$ direction
as the toroidal direction. We set the origin $x=0=y$ exactly 
at the X-point and assume mirror symmetry with respect to the 
$xz$ and~$yz$ planes as well as the translational symmetry 
in the~$z$ direction (see Fig.~\ref{fig-x-point}). We also 
assume steady state.

We are interested in a small vicinity of the X-point, which means 
that we consider locations with $x\ll L$ and $y\ll\delta$. 
In this region the poloidal magnetic field can be generically 
represented by a simple X-point configuration. In terms of the 
poloidal flux function $\Psi(x,y,t)$ this can be written in the 
appropriate gauge as
\beq
\Psi(x,y,t) = - cE_z t + {B_0\delta\over 2}\, 
\biggl( {y^2\over{\delta^2}} - {x^2\over{L^2}} \biggr)\, ,
\label{eq-Psi-Xpoint}
\eeq
with $B_{\rm pol}=\nabla\Psi\times\hat{z}$.

This expression serves as the definition of the scales~$L$ 
and~$\delta$. The first term in this expression is just 
the instantaneous value of the flux at the origin (which 
we can define as the flux that has reconnected since~$t=0$).  
As reconnection proceeds, it increases at a constant rate 
equal to $-c E_z = |cE_z|>0$, where $E_z<0$ is the toroidal
electric field (which is uniform in steady state reconnection). 
The quantity $B_0$ represents the reconnecting magnetic field 
just outside the layer: $B_x(x=0,y=\delta)= B_0$. The poloidal 
magnetic field components corresponding to this flux function are:
\begin {eqnarray}
B_x &=& \partial_y \Psi = B_0\, {y\over\delta} 
\label{eq-Xpoint-Bx} \, ,                               \\
B_y &=& - \partial_x \Psi = B_0\, {\delta\over L} {x\over L}
\label{eq-Xpoint-By} \, . 
\end{eqnarray}

For convenience, we introduce dimensionless variables 
by rescaling $x$, $y$, ${\bf B}$, $\Psi$, and $E_z$ as
\begin {eqnarray}
\bar{x} &\equiv & {x\over L}\, , \qquad \bar{y}\equiv{y\over\delta}\nonumber\\
\bar{\bf B} &\equiv & {{\bf B}\over B_0}\, , \qquad 
\bar{\Psi} \equiv  {\Psi\over{B_0\delta}}\, , \qquad 
\bar{E} \equiv  {cE_z\over{B_0\delta}} \, .
\label{eq-rescaling}
\end{eqnarray}

Then equation~(\ref{eq-Psi-Xpoint}) can be written as
\beq
\bar\Psi' \equiv \bar{\Psi} +\bar{E} t =
\bar{\Psi}(x,y) - \bar{\Psi}(0,0) =
{{\bar{y}^2}\over 2} - {{\bar{x}^2}\over 2} \, .
\label{eq-Psi'}
\eeq

Correspondingly, the shape of a given field line $\bar{\Psi}$
is given (in the two upper quadrants) by 
\beq
\bar{y}(\bar{x},\bar{\Psi}) = 
\sqrt{2\bar{\Psi}' + \bar{x}^2} \, .
\label{eq-shape-yx}
\eeq

Now we want to calculate the motion of the electron fluid.
It is completely determined by two conditions: flux-freezing 
in the poloidal plane (which is not spoiled by the pressure
gradient term in the generalized Ohm law, as we shall discuss 
later) and incompressibility. In order to get an explicit expression 
for the electron velocity, let us consider the trajectory~$\bar{X}(t)$, 
$\bar{Y}(t)$ of an electron fluid element. As it moves through the 
layer, the given fluid element always stays on a field line with 
constant~$\bar{\Psi}$; thus, as one follows its motion, $\bar{\Psi}'=
\bar{\Psi} +\bar{E}t$ varies with time. Correspondingly, the trajectory 
of the element has to satisfy 
\beq
\bar{Y}(t) = \sqrt{2(\bar{\Psi}+\bar{E}t) + \bar{X}^2(t)} \, .
\label{eq-shape-YX}
\eeq

Next, the incompressibility condition implies that the volume per unit flux
following an electron fluid element, $V(\bar{X},\bar{\Psi}')=
V[\bar{X}(t),\bar{\Psi}+\bar{E}t]$, has to be conserved.
Here, $V(\bar{X},\bar{\Psi}')$ is measured along the line~$\bar{\Psi}'$
from the $y$-axis ($\bar{x}=0$) in the case of a field line in the upstream 
region (and from the $x$-axis, $\bar{y}=0$, in the case of a field line in 
the downstream region), up to the fluid element under consideration. 
For example, in the upstream region we thus have
\beq
V[X(t),\Psi'] \equiv 
\int\limits_0^{l(X)}\, {dl\over{|B_{\rm pol}|}}\,\biggl|_{\Psi={\rm const}}=
\int\limits_0^{X(t)}\, {dx\over{|B_x|}}\, \biggl|_{\Psi={\rm const}}=
{\rm const}\,.
\eeq

Using the expression~(\ref{eq-Xpoint-Bx}) for~$B_x$ and~(\ref{eq-shape-yx}) 
for the field line shape~$\bar{y}(\bar{x},\bar{\Psi})$, we get
\beq
V[\bar{X}(t),\bar{\Psi}'] = {L\over B_0}\, 
\int\limits_0^{\bar{X}(t)}\, {d\bar{x}\over{|\bar{y}(\bar{x},\bar{\Psi})|}}
\biggl|_{\Psi={\rm const}} = 
{L\over B_0}\, \log\biggl|{{\bar{X}(t)+\sqrt{\bar{X}^2(t)+2\bar{\Psi}'(t)}}
\over{\sqrt{2\bar{\Psi}'(t)}}}\biggr| = {\rm const} \, .
\label{eq-Xpt-V}
\eeq

Hence, the trajectory is given by
$(\bar{X}(t)+\sqrt{\bar{X}^2(t)+2\bar{\Psi}'})/
\sqrt{2\bar{\Psi}'}={\rm const}$, 
that is $\bar{X}(t)/\sqrt{2\bar{\Psi}'(t)}={\rm const}$. 
Using~(\ref{eq-shape-YX}), we also get a similar expression 
for $\bar{Y}(t)$. Thus,
\begin{eqnarray}
\bar{X}(\bar{X}_0,t) &=& \xi(\bar{X}_0,\bar{\Psi}) \, 
\sqrt{2|\bar{\Psi}+\bar{E}t|} \, ,      \\
\bar{Y}(\bar{X}_0,t) &=& \eta(\bar{X}_0,\bar{\Psi}) \, 
\sqrt{2|\bar{\Psi}+\bar{E}t|} \, ,
\end{eqnarray}
where $\eta= \sqrt{\xi^2\pm 1}$, and where we take ``+'' 
in the upstream region and ``-'' in the downstream region.
The constant parameters~$\xi$ and~$\eta$  represent the 
initial position of the electron fluid element at $t=0$.

Differentiating these expressions with respect to time and using 
$d\bar{\Psi}'/dt=\bar{E}$, we obtain the electron velocity field 
at any point $(\bar{x},\bar{y})$:
\begin{eqnarray}
\bar{v}_{\bar{x}}^{(e)}(\bar{x},\bar{y}) &=& {\bar{x}\over 2} \, 
{\bar{E}\over{\bar{\Psi}'(\bar{x},\bar{y})}} = 
- \bar{x}\, {|\bar{E}|\over{\bar{y}^2-\bar{x}^2}} \, ,\\
\bar{v}_{\bar{y}}^{(e)}(\bar{x},\bar{y}) &=& {\bar{y}\over 2} \, 
{\bar{E}\over{\bar{\Psi}'(\bar{x},\bar{y})}} = 
- \bar{y}\, {|\bar{E}|\over{\bar{y}^2-\bar{x}^2}} \, .
\end{eqnarray}

Notice that $\bar{v}_{\bar{y}}^{(e)}/\bar{v}_{\bar{x}}^{(e)}=\bar{y}/\bar{x}$, 
so electrons flow along purely radial lines, inward in the upstream region 
and then outward in the downstream region. Even though all the streamlines 
converge to the X-point in the upstream region (and fan out of the X-point 
in the downstream region) the motion is incompressible since the magnitude 
of the velocity diverges near the origin. Also note that if the above velocity 
field holds, electrons never actually cross the separatrix; they all go 
through the X-point. This is of course an artifact of our ideal eMHD 
assumption.

This poloidal electron velocity field results in a poloidal electric current:
\begin{eqnarray}
j_x^{(e)} &=& -\, e n_e v_x^{e} = -\, e n_e L \bar{v}_{\bar{x}}^{e} 
\label{eq-Xpoint-jx}              \, ,              \\
j_y^{(e)} &=& -\, e n_e v_y^{e} = -\, e n_e \delta \bar{v}_{\bar{y}}^{e} 
\label{eq-Xpoint-jy}             \, .
\end{eqnarray}
This current in turn produces a toroidal magnetic field.
According to Ampere's law we have
\begin{eqnarray}
\partial_y B_z^{(e)} &=& {4\pi\over c}\, j_x^{(e)} = 
{{4\pi n_e e}\over c}\, L |\bar{E}| \, 
{\bar{x}\over{2\bar{\Psi}'(\bar{x},\bar{y})}}          \, ,  \\
\partial_x B_z^{(e)} &=& -\, {4\pi\over c}\, j_y^{(e)} = 
-{{4\pi n_e e}\over c}\, \delta |\bar{E}| \, 
{\bar{y}\over{2\bar{\Psi}'(\bar{x},\bar{y})}} \, .
\end{eqnarray}

We can thus compute the resulting toroidal field by integrating either 
of these equations while taking into account that $B_z(0,y)=0=B_z(x,0)$ 
because of symmetry. In the upstream region it is convenient to integrate
$\partial_x B_z^{(e)}(x,y)$ with respect to~$x$ at constant~$y$ starting 
with $x=0$. (In the downstream region, it is convenient to do the opposite.) 
Thus we can write:
\beq
B_z(\bar{x},\bar{y}>\bar{x})= \int\limits_0^{\bar{x}} 
L\partial_x B_z d\bar{x}= 
-{{4\pi n_e e}\over c}\, \delta L |\bar{E}| \bar{y} \, 
\int \limits_0^{\bar{x}} {d\bar{x}\over{2\bar{\Psi}'(\bar{x},\bar{y})}}\, .
\label{eq-Bz-1}
\eeq
Using the expression~(\ref{eq-Psi'}) for $\bar{\Psi}'(\bar{x},\bar{y})$, 
we get 
\beq
B_z(\bar{x},\bar{y})= -\, {1\over 2}\, B_0 Q\, 
\log\vert{{\bar{y}+\bar{x}}\over{\bar{y}-\bar{x}}}\vert \, ,
\label{eq-Bz-2}
\eeq
where we define a dimensionless constant 
\beq
Q \equiv {{4\pi n_e e}\over{cB_0}}\, \delta L |\bar{E}| =
{\delta\over d_i} {L\over V_A} |\bar{E}| \, .
\label{eq-def-Q}
\eeq
Expression~(\ref{eq-Bz-2}) is actually valid in both the upstream and 
downstream regions. We note that a very similar expression was obtained, 
in the eMHD framework, in Ref.~\cite{Biskamp-1997}.

We can rewrite the coefficient $Q$ in a different form by expressing 
the reconnection electric field~$\bar{E}$ in terms of the other parameters 
of the reconnection region. Indeed, let $v_{\rm rec}$ be the reconnection 
velocity, $v_{\rm rec}=-v_y(x=0,y\gg\delta)$. Then $|E_z|=v_{\rm rec}B_0/c$ 
and hence $|\bar{E}|=v_{\rm rec}/\delta$. Then, we get
\beq
Q = {L\over d_i}\, {v_{\rm rec}\over V_A} = 
C\, {\delta\over d_i}\, {u\over V_A} \, ,
\label{eq-Q}
\eeq
where $u$ is the velocity of the flow out of the layer, and where we
define a new dimensionless parameter~$C$:
\beq
C\equiv {{L v_{\rm rec}}\over{\delta u}} = O(1) \, .
\label{eq-def-C}
\eeq
Because of the condition of overall mass conservation we expect~$C$ 
to be of order unity.

Usually, one expects the thickness~$\delta$ of a Hall-MHD reconnection 
region to be comparable to the ion collisionless skin depth~$d_i$ and 
the outflow velocity~$u$ to be of order~$V_A$. Thus, equation~(\ref{eq-Q})
tells us that the proportionality coefficient~$Q$ is, generally speaking,
expected to be of order unity. In practice, however, $Q$ may significantly 
deviate from unity for any specific physical system. For example, in the 
MRX experiment one often encounters $\delta\simeq d_i/3$ and~$u<V_A$, 
so~$Q$ can be smaller than~$1/3$.

The reader should be warned that formula~(\ref{eq-Bz-2}) only applies to 
our specific example for the poloidal field, eq.~(\ref{eq-Psi-Xpoint}), 
and may not be applicable to various configurations realized in some
numerical simulations and in the~MRX. We chose this specific example 
because of its simplicity and clarity and we leave more complicated 
poloidal field structures for a future study. We believe that the most 
physically-relevant among these other structures is a configuration 
with an inner electron current sheet (electron dissipation region) 
of a finite width in the $x$-direction.

As we see from equation~(\ref{eq-Bz-2}), in the simple X-point configuration
considered in this section, the assumptions of ideal eMHD lead to a logarithmic
divergence of~$B_z$ at the separatrix~$\bar{y}=\bar{x}$. Later, in~Sec.~\ref
{subsec-inertia} we shall discuss how this singularity is removed by including 
finite electron inertia. We also see that $B_z$ in our solution is constant 
along straight radial rays $y={\rm const}\cdot x$. 
In reality we expect electron inertia and other non-ideal effects to intervene 
and break this idealized picture near the separatrix. However, we believe that 
the main tendency for the constant-$B_z$ contours to be strongly elongated 
along the separatrix will survive. In fact, this overall behavior is in a 
very good agreement with the results of numerical simulations~\cite
{Biskamp-1997,Shay-1998a,Wang-2000,Arzner-2001,Pritchett-2001,Breslau-2003}, 
and is also consistent with the experimental data~\cite{Ren-2005}.


\section{Stationary Ideal Incompressible eMHD in 2.5 Dimensions:
General Results}
\label{sec-general}

In this section we step back from the particular
example of the previous section and derive several
general results that are valid in steady-state 
2D incompressible eMHD for an arbitrary poloidal 
field structure. 


\subsection{General relationships between toroidal magnetic 
field, electron stream function and the volume per flux in eMHD}
\label{subsec-Bz=Phi=V}

First we introduce three important functions:
the volume-per-flux integral~$V(x,\Psi)$, the electron stream function
$\Phi_e$ and the (electron contribution to) the toroidal magnetic field~$B_z$.
We derive important relationships between these functions and discuss 
their implications.

The volume-per-flux integral~$V(x,\Psi)$ is defined (in the upstream region), 
as in the last section, by
\beq
V(x,\Psi) \equiv \int\limits_0^x \ {{dl}\over{|B|}}\biggl|_\Psi = 
\int\limits_0^x \ {{dl_{\rm pol}}\over{|B_{\rm pol}|}}\biggl|_\Psi \, .
\label{eq-def-V}
\eeq
where, in the last expression, the integration is performed along a given 
poloidal line, $\Psi$, from the $y$-axis $x=0$ to the given point~$(x,\Psi)$.
A similar expression can be defined in the downstream region.

The electron stream function~$\Phi_e(x,y)$ is defined, for an incompressible
flow,  $\nabla\cdot{\bf v}^{(e)}=0$, by the poloidal electron velocity as 
\beq
{\bf v}_{\rm pol}^{(e)} = [\nabla\times(\Phi_e\hat{z})] = 
[\nabla\Phi_e\times\hat{z}]    \, .
\label{eq-def-Phi_e}
\eeq
For definiteness, we choose the streamline corresponding to 
zero~$\Phi_e$ to coincide with the line from which we count the 
volume-per-flux, i.e., with the $y$-axis: $\Phi_e(0,y)\equiv~0$.
 
With these definitions we now show that in steady state 
ideal incompressible electron MHD the two functions are 
just proportional to each other:
\beq
\Phi_e(x,y) = -\, c E_z V (x,y) \, .
\label{eq-Phi_e=V}
\eeq
where $E_z$ is the uniform electric field in the $z$-direction.

To prove this, let us consider the variation of~$\Phi_e$ {\it along the 
poloidal magnetic field} and show that it is proportional to that of~$V$. 
Consider two points lying close to each other on the same poloidal field 
line~$\Psi$, and let $\Delta l_{\rm pol}$ be the infinitesimal separation 
between these two points along the poloidal field. From equation~(\ref
{eq-def-V}), the difference between the volume-per-flux of these two 
points is
\beq
\Delta V = {\Delta l_{\rm pol}\over{B_{\rm pol}}} \, .
\label{eq-DeltaV-paralell}
\eeq

On the other hand, the difference between the values of the electron stream 
function at these two points can be expressed in terms of the perpendicular 
component of the poloidal electron velocity 
\beq
{\bf v}_{\rm pol,\perp}^{(e)} = [\nabla\Phi_e \times \hat{z}]_\perp = 
{{\Delta\Phi_e}\over{\Delta l_{\rm pol}}}\ [{\bf b}_{\rm pol}\times\hat{z}]=
-\, {{\Delta\Phi_e}\over{\Delta l_{\rm pol}B_{\rm pol}}}\, \nabla\Psi \, , 
\label{eq-v_pol_perp-1}
\eeq
where ${\bf b}_{\rm pol}$ is the unit vector along the poloidal magnetic 
field. On the other hand, ${\bf v}_{\rm pol,\perp}^{(e)}$ can be deduced 
from the generalized Ohm law. Neglecting resistive and inertial terms and 
taking into account that $\partial_z p_e = 0$, the toroidal component of 
this law can be written as
\beq
c\, E_z\hat{z}+[{\bf v}_{\rm pol,\perp}^{(e)}\times{\bf B}_{\rm pol}] = 0 \, .
\label{eq-Ohm-z}
\eeq
By taking the vector product of this equation with ${\bf B}_{\rm pol}$, 
we get 
\beq
{\bf v}_{\rm pol,\perp}^{(e)} = 
c\, {{E_z\hat{z}\times {\bf B}_{\rm pol}}\over{B_{\rm pol}^2}} = 
c\, {E_z\over{B_{\rm pol}}}\, [\hat{z}\times{\bf b}_{\rm pol}] = 
c\, {E_z\over{B_{\rm pol}^2}}\, \nabla\Psi \, .
\label{eq-v_pol_perp-2}
\eeq
Comparing equations~(\ref{eq-v_pol_perp-1}) and~(\ref{eq-v_pol_perp-2}) 
we immediately see that
\beq
\Delta\Phi_e = -\, {cE_z\over{B_{\rm pol}}}\, \Delta l_{\rm pol} \, ,
\eeq
which, together with equation~(\ref{eq-DeltaV-paralell}), gives
\beq
\Delta\Phi_e = -\, cE_z \Delta V  \, .
\eeq

Because we choose to count the volume-per-flux from the $\Phi_e=0$ 
electron streamline, then, summing this equation along each field line, 
we get equation~(\ref{eq-Phi_e=V}), as desired. But we can actually 
proceed more directly. Let us consider the variations of~$\Phi_e$ and~$V$ 
{\it across the poloidal field}. For this, consider a single poloidal 
field line~$\Psi$ at two neighboring moments of time, $t_1$ and 
$t_2=t_1+\Delta t$. Consider a certain point $A_1$ on the 
field line at $t=t_1$ and see it ${\bf E\times B}$-drift 
with the field line to a new location $A_2$ at $t=t_2$. 
The corresponding change in the volume-per-flux is due 
to the plasma that has flowed in along the poloidal field 
past this point during the time~$\Delta t$:
\beq
\Delta V \equiv V(A_2)-V(A_1) = 
-\, {{v_{\rm pol,\parallel} \Delta t}\over{B_{\rm pol}}} \, .
\label{eq-DeltaV-perp}
\eeq
At the same time, this parallel inflow of the plasma results in
a change in~$\Phi_e$ between points~$A_1$ and~$A_2$:
\beq
\Delta \Phi_e \equiv \Phi_e (A_2) - \Phi_e (A_1) = v_{\rm pol,\parallel}\, 
(\Delta{\bf s}_{\rm pol,\perp}\cdot[\hat{z}\times{\bf b}_{\rm pol}]) \, ,
\eeq
where the poloidal displacement vector $\Delta{\bf s}_{\rm pol,\perp}$ 
in the direction perpendicular to the poloidal magnetic field is given 
by the ${\bf E\times B}$ drift:
\beq
\Delta{\bf s}_{\rm pol,\perp} = {\bf v}_{\perp,\rm pol} \Delta t = 
\Delta t\, {cE_z\over{B_{\rm pol}}}\, [\hat{z}\times{\bf b}_{\rm pol}] \, .
\eeq
Thus, we get
\beq
\Delta \Phi_e = v_{\rm pol,\parallel}\Delta t\, {cE_z\over{B_{\rm pol}}}\, ,
\eeq
and comparing this result with equation~(\ref{eq-DeltaV-perp}),
we again see that 
\beq
\Delta\Phi_e = -\, cE_z \Delta V  \, .
\eeq
A similar derivation also holds in the downstream region.

Thus we have shown that the variation of the electron stream function~$\Phi_e$
in both parallel and perpendicular directions is equal to $-cE_z$ times the 
corresponding variation in the volume per flux integral~$V$. Using the 
convention of counting both~$\Phi_e$ and~$V$ starting from the y-axis, 
we again arrive at the relationship~(\ref{eq-Phi_e=V}).

The second important relationship is the proportionality between 
the electron stream function~$\Phi_e$ and the electron contribution 
to the toroidal field~$B_z$. This well-known relationship follows 
immediately from Ampere's law and the reflection symmetry conditions 
[$B_z(x=0,y)=0=\Phi_e(x=0,y)$ upstream and~$B_z(x,y=0)=0=\Phi_e(x,y=0)$ 
downstream]. It reads: 
\beq
B_z(x,y) = -\, D\, \Phi_e(x,y) \, ,
\label{eq-B_z=Phi_e}
\eeq
with the coefficient $D$ given by 
\beq
D \equiv {{4\pi n_e e}\over c} = {\sqrt{4\pi\rho}\over{d_i}} =
{B_0\over{d_i V_A}} \, .
\label{eq-def-D}
\eeq
(Here $B_0$ is an arbitrary normalization field used in the definition 
of~$V_A$.)

Note that the coefficient~$D$ defined by equation~(\ref{eq-def-D})
is constant for the case of uniform density considered here. 
If the density were not uniform, we would get a similar result 
$B_z\sim\Phi_e$, if the electron density is incorporated into~$\Phi_e$, 
i.e., if $\Phi_e$ is defined 
by $n_e{\bf v}_{\rm pol}^{(e)}=[\nabla\times(\Phi_e\hat{z})]$.
Similarly, equation~(\ref{eq-Phi_e=V}) would be valid if~$V$
is understood as the number of electrons per unit flux, 
instead of the volume-per-flux.

Combining this equations~(\ref{eq-B_z=Phi_e}) and~(\ref{eq-Phi_e=V}) 
we immediately see that
\beq
B_z = cD E_z V \, .
\label{eq-B_z=V}
\eeq
This result is important because it shows how, 
in eMHD, one can immediately determine~$B_z$ once 
the poloidal field structure is known, without having 
to solve any partial differential equations! One just 
has to compute the volume-per-flux integral given by
equation~(\ref{eq-def-V}) upstream and the corresponding 
expression downstream, and the toroidal field will be just 
the constant~$cDE_z$ times it.

Already by itself, the simple relationship~(\ref{eq-B_z=Phi_e}) 
is important, because it means that, neglecting ion currents, 
the toroidal field is constant along the poloidal electron 
streamlines; its value is simply transported in space by 
the poloidal electron flow: 
\beq
({\bf v}_{\rm pol}^{(e)}\cdot\nabla) B_z =
-\, D B_0\, [\nabla\Phi_e\times\hat{z}]\cdot\nabla\Phi_e \equiv 0 \, ,
\label{eq-tor-magn-induction}
\eeq
where ${\bf v}_{\rm pol}^{(e)}$ is the total poloidal electron velocity
including the parallel flow.
 
This result suggests that the toroidal magnetic field cannot be created 
locally, in the inner part of the reconnection region (where ion current 
is unimportant); instead, it has to be brought into this region by the 
convergent electron flow. The toroidal field thus has to be generated 
in the outer parts of the layer, where the ion-current contribution 
to~$B_z$ is not negligible. We discuss this generation process 
in~Sec.~\ref{subsec-ion-currents}.

In addition, as long as electrons are completely frozen 
into the magnetic field, the evolution equation for the 
toroidal magnetic field, i.e., the toroidal component of 
the magnetic induction equation, tells us that 
\beq
v_{\rm pol}^{(e)}\cdot\nabla B_z = 
B_{\rm pol}\cdot\nabla v_z^{(e)} \, ,
\label{eq-drake}
\eeq
in a steady state. Therefore, since the left-hand 
side (LHS) of this equality is zero, as we have just 
shown, the right-hand side (RHS) is also zero. 
That is, the toroidal electron velocity
and hence the toroidal current density $j_z=-\,en_e v_z^{(e)}$
are uniform along poloidal field lines as long as one can
neglect the ion current. Since from Ampere's law the toroidal 
current density is simply proportional to the Laplacian of~$\Psi$, 
we see that in order to get a consistent solution, one cannot pick 
the poloidal flux function arbitrarily; one has to impose the
condition that $\nabla^2\Psi = F(\Psi)$. For example, the simple 
X-point poloidal field configuration considered in Sec.~\ref{subsec-Xpoint} 
trivially satisfies this condition with $F(\Psi)\equiv{\rm const}$.


The above observation also means that within the pure eMHD framework
with no ion currents one cannot really apply the well-known conventional 
explanation, first introduced in Ref.~\cite{Mandt-1994}, of how 
the quadrupole field is generated. 
By itself, this argument does not rely on neglecting ion currents.
Instead it relies on the fact that magnetic field is completely 
frozen into the electron fluid. The toroidal field is then viewed 
as being produced from the poloidal magnetic field as a result 
of the differential stretching in the toroidal direction by the 
electron flow.
This argument thus approaches the generation of the quadrupole field
from a different angle: it presents the point of view of the ideal 
eMHD Ohm's law (with the Hall term), instead of using Ampere's law. 
It basically goes like this: as a field line is advected into the 
reconnection layer, the electrons on it start to move toroidally 
(to carry some of the reconnection current) and they do it 
differentially, moving faster on the central piece of the 
field line. Since the magnetic field is frozen into the electron 
fluid, the field line bends out of the reconnection plane, resulting
in the quadrupole pattern of the toroidal field. This line of thought 
is actually quantified by equation~(\ref{eq-tor-magn-induction}).
That is, the toroidal field is created by the stretching due to 
the non-uniformity of $v_z^{(e)}$ along a poloidal field line (RHS) 
and is advected with the poloidal electron flow (LHS). But, as we 
have just seen, as long as one neglects ion currents, this equation 
becomes simply $0=0$. Thus, in order to understand toroidal field 
generation, one first needs to take ion currents into account 
(see~Sec.~\ref{subsec-ion-currents} for more discussion).


\subsection{The Shape of Field Lines in the $xz$ Plane}
\label{subsec-shape-xz}

A complimentary way to look at the problem of the toroidal
field generation is to analyze the shape of a field line 
projected on the $xz$~plane, and see how it changes as the 
field lines move deeper into the layer. Let this shape be 
represented by the function~$z(x,\Psi)$, which is given by
\beq
{dz\over dx}\vert_{\Psi} = {B_z\over B_x} \, .
\eeq
By integrating this along a field line we obtain:
\beq
\Delta z(x,\Psi) = z(x,\Psi)-z(0,\Psi) = \int\limits_0^x\, B_z(x,\Psi)
{dx\over{B_x(x,\Psi)}} \vert_{\Psi={\rm const}}   \, .
\eeq

Using equations~(\ref{eq-B_z=V}) and~(\ref{eq-def-V}), we then have
\beq
\Delta z(x,\Psi) =  cDE_z\, {{V^2(x,\Psi)}\over{2}}   \, .
\label{eq-shape-zx}
\eeq

Note that the volume per flux is conserved by the motion of the electron 
fluid; this means that the $x$-position of an electron fluid element that 
stays on some constant-$\Psi$ field line changes with time in such a way 
as to keep $V(x,\Psi)$ constant. Thus, if one follows a specific fluid 
element on a given moving field line~$\Psi$, one finds that the toroidal 
distance~$\Delta z(x,\Psi)$ between this element and the point where the 
field line intersects the $x=0$~plane does not change with time. On the 
other hand, as we showed in~Sec.~\ref{subsec-idea}, the fluid element 
moves along the {\it poloidal} magnetic field towards the $y$~axis, and 
so its $x$-coordinate decreases. Therefore, the shape of the field line,
which can be characterized by the function $x(\Delta z)$, changes with 
time. Interestingly, this change is not due to the differential toroidal 
stretching, as it is usually assumed, but is simply due to the fact that 
the two mirror-symmetric parts of the line are squeezed together by the 
converging poloidal flow.


\subsection{The Role of Ion Currents in the Generation of the
Quadrupole Field}
\label{subsec-ion-currents}

As we saw in Sec.~\ref{subsec-Bz=Phi=V}, when the current due to 
the ions is neglected, the toroidal field is conserved along the 
electron streamlines and hence cannot be locally generated in 
the inner part of the reconnection layer. This indicates that, 
in order to explain how and where the toroidal field is generated, 
one has to bring the ions back into the picture. Deep inside the 
reconnection region, at $x\ll L$ and~$y\ll\delta\sim d_i$, 
the poloidal ion currents are indeed negligible and the above 
picture applies. On the other hand, in the upstream region well 
outside of the reconnection layer (i.e., for $y\gg\delta$), 
ideal one-fluid MHD works well.
In this region the electrons do move poloidally towards the reconnection 
layer with the ${\bf E\times B}$ velocity and the associated electron current 
would generate the toroidal field; however, the ions also move happily
in the same direction and with the same speed. As a result, the 
ion-current contribution to the toroidal field exactly cancels 
that of the electrons. Thus, the net toroidal field is zero 
in this region. From this we see that, in order to understand where the 
quadrupole toroidal field comes from, one has to look at the 
outskirts of the reconnection layer, where the ions become
partly decoupled from the electrons, so that $0<|j_y^{(i)}|
<|j_y^{(e)}|$.

We can discuss the toroidal field generation from a different point of view, 
in terms of the shape $z(x,\Psi)$ of a given field line~$\Psi$ 
as it is carried into the current layer by the electron flow. 
Far upstream, this field line lies entirely in the reconnection ($x,y$) 
plane, but as it moves into the reconnection region, it gradually starts
to bend out of this plane. The toroidal electron velocity can
be non-uniform along the line only in this transition region 
of non-zero ion current. Correspondingly, toroidal field 
is produced inside this region; subsequently, deeper inside 
the reconnection layer, the toroidal electron velocity becomes uniform along
the line and hence the toroidal elongation freezes. Any further 
lengthening of the field line in the toroidal direction can be 
directly attributed to the ``injection'' of new segments of the 
field line in the transition region.


\subsection{Toroidal Electron Velocity}
\label{subsec-v_z}

To illustrate this picture, let us consider an extremely 
simplified model where the transition region is a razor-thin
line $y=\delta$. In this example, the electrons and ions move 
together above $y=\delta$ and so $B_z(x,y>\delta)\equiv 0$.
Below this sharp boundary, we shall regard the ions as poloidally 
motionless, so that $j_{\rm pol}^{(i)}(y<\delta)=0$ and hence the 
pure eMHD picture developed in the preceding sections applies. 
In addition, in this and in the next section we shall, 
for simplicity, neglect the toroidal component of the 
diamagnetic electron flow that results from poloidal
electron pressure gradient.

Now let us consider a given field line~$\Psi$; as we follow its 
motion through the layer, the magnetic flux~$\Psi'$ between the 
separatrix and the given field line changes linearly in time 
according to $\Psi'(t)=\Psi+cE_zt$. Denote the $x$-coordinate 
of the point where this field line intersects the boundary 
$y=\delta$ by $x_\delta[\Psi'(t)]=x_\delta(\Psi+cE_zt)$. 
For example, in the simple X-point configuration considered 
in Sec.~\ref{subsec-Xpoint}, we have 
$\bar{x}_\delta=[1-2\bar{\Psi}'(t)]^{1/2}$. 
Next, because there is no toroidal field above $y=\delta$, we can 
set $z=0$ everywhere along this boundary, i.e., $z[x_\delta(\Psi'),\Psi]=0$. 
Then, using equation~(\ref{eq-shape-zx}), we can express the toroidal 
coordinate of any fluid element $(X(t),\Psi)$ on a given line~$\Psi$  
as
\beq
z[X(t),\Psi]=\Delta z[X(t),\Psi] - \Delta z[x_\delta(\Psi'),\Psi] =
cDE_z \, {{V^2[X(t),\Psi]-V_\delta^2(\Psi')}\over 2}   \, ,
\label{eq-z_x_Psi}
\eeq
where 
\beq
V_\delta(\Psi') \equiv V[x_\delta(\Psi'(t)),\Psi] \, .
\label{eq-def-V_delta}
\eeq

In this section we are interested in the toroidal electron velocity,
so let us see how $z(X,\Psi)$ changes with time following an electron 
fluid element. To do this, differentiate equation~(\ref{eq-z_x_Psi}) 
with respect to time. When doing this we have to take into account 
that in ideal incompressible electron MHD the motion~$X(t)$ of a given 
fluid element is constrained by the condition that $V[X(t),\Psi]$ remains 
constant. Then we have
\beq
v_z^{(e)}[X(t),\Psi] = {d\over{dt}}\, z[X(t),\Psi] = 
-\, cDE_z\, {d\over{dt}}\, \biggl[{{V_\delta^2(\Psi')}\over 2}\biggr] =
-\,c^2 DE_z^2\,{d\over{d\Psi'}}\,\biggl[{{V_\delta^2(\Psi')}\over 2}\biggr]\,,
\label{eq-v_z}
\eeq
since $d\Psi'/dt=cE_z$. Thus, the velocity is proportional to 
the flux derivative of the square of a flux tube's entire volume 
up to the boundary~$y=\delta$. 

One sees that the toroidal velocity is constant along field lines but, 
in general, varies from line to line. In particular, the volume-per-flux
$V_\delta(\Psi')$ grows rapidly near the separatrix and so $v_z^{(e)}$
becomes very large there. This appears to be inconsistent, for instance,
with the simple X-point configuration considered in~Sec.~\ref{subsec-Xpoint};
indeed, the particular form of the poloidal flux function in that example
corresponded to a flat toroidal current profile, $j_z={\rm const}$, and 
hence~$v_z^{(e)}={\rm const}$. The way to resolve this discrepancy is to
note that the sharp rise in the toroidal current density that corresponds 
to equation~(\ref{eq-v_z}), leads to only a relatively small change in 
the poloidal field structure. Moreover, this change is actually consistent 
with that expected from the back-reaction of the toroidal magnetic field 
pressure. This back-reaction arises because, as one approaches the separatrix,
the toroidal field increases sharply and starts to play an important dynamical 
role. In particular, it modifies the poloidal field structure through the
vertical pressure balance condition; the poloidal field decreases near the 
separatrix and this leads (by Ampere's law) to an additional electric current, 
strongly concentrated near the separatrix. We can estimate this additional
electric current as follows.

Let us write the vertical pressure balance as 
\beq
B_{\rm pol}^2 +B_z^2 = B_0^2 - 8\pi P = B_0^2 \bar{y}^2 \, ,
\label{eq-pressure-balance}
\eeq
where we have assumed that the total plasma pressure~$P$ has
a parabolic profile: $8\pi P=B_0^2(1-\bar{y}^2)$. If we neglect
the toroidal field pressure term in this equation, we then recover
our original poloidal field profile $B_{\rm pol}\approx B_x=B_0\bar{y}$,
which corresponds to uniform toroidal current. Note, however, that
even if $B_x$ itself is small, its rate of change may become important 
near the separatrix, so that the corresponding small but rapid change 
in~$B_x$ results in a large contribution to the toroidal current.
Indeed, differentiating equation~(\ref{eq-pressure-balance}) with 
respect to~$\bar{y}$, we get
\beq
{{dB_{\rm pol}^2}\over{d\bar{y}}} = 2 B_x {dB_x\over{d\bar{y}}} 
\simeq -\, {8\pi\over c}B_x\delta\, j_z = 
2B_0^2 \bar{y} - {{dB_z^2}\over{d\bar{y}}}     \, .
\eeq
Then, using equation~(\ref{eq-Xpoint-Bx}) and~(\ref{eq-B_z=V}), we find
\beq
j_z = -\, {c\over{4\pi}}\, \biggl[{B_0\over\delta} - c^2 D^2 E_z^2\, 
{\partial\over{\partial\Psi'}}\, \biggl({V^2(x,\Psi')\over 2}\biggr)\biggr]
\simeq n_e e c^2 D E_z^2\, 
{\partial\over{\partial\Psi'}}\, \biggl[{V^2(x,\Psi')\over 2}\biggr]\, ,
\label{eq-j_z}
\eeq
and so, assuming that the additional toroidal current is predominantly 
carried by electrons,
\beq
v_z^{(e)} \simeq -\, c^2 D E_z^2\, 
{\partial\over{\partial\Psi'}}\, \biggl[{V^2(x,\Psi')\over 2}\biggr]
\eeq
--- an expression that is very similar, although not quite the same, 
as equation~(\ref{eq-v_z}).

Finally, for reference, let us give expressions for~$V_\delta(\Psi')$ 
and~$v_z^{(e)}$ that correspond to the simple X-point configuration 
considered in~Sec.~\ref{subsec-Xpoint}. 
First, according to equation~(\ref{eq-Xpt-V}), we have
\beq
V_\delta(\bar{\Psi}') = {L\over B_0}\, 
\log\,\biggl|\,{{1+\sqrt{1-2\bar{\Psi}'}}\over{\sqrt{2\bar{\Psi}'}}}\,\biggr|\,
= {L\over{2B_0}}\, \log\, \biggl|\, 
{{1+\bar{x}_\delta(\Psi')}\over{1-\bar{x}_\delta(\Psi')}}\,\biggr| \, ,
\label{eq-Xpt-V_delta}
\eeq
so that $(d/d\bar{\Psi}')\,V_\delta(\bar{\Psi}')=
-(L/2B_0\bar{\Psi}'\bar{x}_\delta(\Psi')$.
 
Then, from equation~(\ref{eq-v_z}) we obtain
\beq
v_z^{(e)}(\bar{\Psi}') = -\, B_0\delta D |\bar{E}|^2\, 
V_\delta(\bar{\Psi}')\, {d\over{d\bar{\Psi}'}}\, V_\delta(\bar{\Psi}') =
{{L^2\delta D}\over{2B_0}}\,  
{{|\bar{E}|^2}\over{2\bar{\Psi}'\bar{x}_\delta(\bar{\Psi}')}}\,
\log\biggl|{{1+\bar{x}_\delta(\bar{\Psi}')}\over{1-\bar{x}_\delta(\bar{\Psi}')}}\biggr|\,.
\label{eq-Xpt-v_z-1}
\eeq
Finally, using definition~(\ref{eq-def-Q}), we can write this as
\beq
v_z^{(e)}(\bar{\Psi}') = 
{{LQ|\bar{E}|}\over{4\bar{\Psi}'\bar{x}_\delta(\bar{\Psi}')}}\,
\log\biggl|{{1+\bar{x}_\delta(\bar{\Psi}')}\over{1-\bar{x}_\delta(\bar{\Psi}')}}\biggr|\,.
\label{eq-Xpt-v_z-2}
\eeq


\subsection{Finite Electron-Inertia Effects}
\label{subsec-inertia}

We can use the above formula for $v_z^{(e)}$ to estimate 
when the electron inertial term stops being negligible in
the toroidal component of the electron equation of motion.
From that moment on, the finite electron inertia will be 
large enough to balance part of the toroidal electric field, 
and thus, the electrons will no longer have a pure ${\bf E\times B}$ 
velocity and will no longer follow the field lines exactly. 
The inertial term (for a single electron) can be written as
\beq
m_e ({\bf v}\cdot \nabla)\, v_z^{(e)} \approx
m_e ({\bf v}_{\rm pol,\perp} \cdot \nabla) \, v_z^{(e)} =
m_e ({\bf v}_{\rm pol,\perp} \cdot \nabla\Psi')\, 
{{dv_z^{(e)}}\over{d\Psi'}} \, ,
\eeq
where we take into account that $v_z^{(e)}$ is constant along field lines
and so is a function of~$\Psi'$ only. Using expression~(\ref{eq-v_pol_perp-2})
for ${\bf v}_{\rm pol,\perp}$, we get
\beq
m_e({\bf v}\cdot\nabla)\, v_z^{(e)} \approx 
m_e cE_z\,{{dv_z^{(e)}}\over{d\Psi'}}\, .
\label{eq-v-dot-del-v_z}
\eeq
Then, using~(\ref{eq-v_z}), we get
\beq
m_e ({\bf v}\cdot \nabla)\, v_z^{(e)} \approx -\, m_e c^3 E_z^3 D \,
{d^2\over{d(\Psi')^2}}\, \biggl[{{V_\delta^2(\Psi')}\over 2}\biggr] \, .
\label{eq-inertia}
\eeq

We can no longer neglect electron inertia when this term becomes 
comparable to the toroidal electric force on an electron, $-\,e E_z$. 
We estimate this to happen for values of~$\Psi'$ of order of
$\Psi_*^\prime$, which is obtained as the solution of the equation
\beq
\bigl[V_\delta^2(\Psi')\bigr]^{\prime\prime} = 
{2e\over{m_e cD}}\, {1\over{c^2 E_z^2}} = 
{1\over{2\pi n_e m_e}}\, {1\over{c^2 E_z^2}} \, .
\eeq

We can apply this estimate to our simple X-point example,
for which $V_\delta(\Psi')$ is given by~(\ref{eq-Xpt-V_delta}).
Since we expect the electron inertia to become important
only near the separatrix, $\bar{\Psi}'\ll 1$, we can approximately
write $V_\delta(\Psi')\simeq - (L/2B_0) \log{\bar{\Psi}'}$,
and then
\beq
v_z^{(e)} \approx -\, {1\over 2}\ QL|\bar{E}|\ 
{\log{\bar{\Psi}'}\over{2\bar{\Psi}'}} \, .
\eeq
The inertial term in the toroidal component of Ohm's law is then estimated, 
with the help of equation~(\ref{eq-v-dot-del-v_z}), as 
\beq
m_e ({\bf v}\cdot \nabla)\, v_z^{(e)} \approx 
m_e \bar{E}\, {{dv_z^{(e)}}\over{d\bar{\Psi}'}} \approx
-\, m_e Q L |\bar{E}|^2\, {{\log{\bar{\Psi}'}}\over{4\bar{\Psi}'^2}} \, ,
\eeq
where we recall that $\bar{E}<0$ in our solution.
After some manipulation we can write the condition on $\bar{\Psi}'_*$ as
\beq
{{2\bar{\Psi}'_*}\over{\sqrt{|\log{\bar{\Psi}'_*}|}}}= Q\ {d_e\over{\delta}}= 
\sqrt{m_e\over{m_i}}\ {{Lv_{\rm rec}}\over{\delta V_A}} = 
O\biggr[\biggl({m_e\over{m_i}}\,\biggr)^{1/2}\biggr]  \, ,
\eeq
where we use equation~(\ref{eq-Q}) and define $d_e\equiv c/\omega_{pe}$.
Thus,
\beq
\bar{\Psi}'_* \sim \sqrt{{m_e\over{m_i}}\, \log{m_i\over{m_e}}} \, .
\label{eq-Psi_*}
\eeq

When traced to the $y$~axis, this critical field line corresponds to
a distance 
\beq
y_* \sim \biggl({m_e\over{m_i}}\,\log{m_i\over{m_e}}\biggr)^{1/4}\ \delta
\label{eq-y_*}
\eeq
from the X-point. (This is of order $\delta/4$ for hydrogen plasma).
It is essentially (apart from the logarithmic factor) of the same 
order as the distance at which electrons become demagnetized, i.e., 
comparable to the size of electron figure-eight and betatron orbits.


\subsection{The bipolar poloidal electric field}
\label{subsec-bipolar-E}

Why do field lines move in the toroidal direction
as they enter the layer? To answer this question, we need 
to consider the toroidal projection of the perpendicular
(to the total magnetic field) electron velocity, $v_{\perp,z}$. 
Let us  locally introduce a rotated orthonormal coordinate 
system $(x',y',z)$ where~$x'$ is the direction along the 
poloidal magnetic field and~$y'$ is the direction in the 
poloidal plane which is perpendicular to the poloidal magnetic 
field. Taking into account the electron pressure (which we 
assume isotropic) but neglecting the electron inertia, 
we can express $v_{\perp,z}^{(e)}$ as
\begin{eqnarray}
v_{\perp,z}^{(e)} &=& 
c\,{{[{\bf E}_{\rm pol}\times{\bf B}_{\rm pol}]_z}\over{B^2}}
\,+\,c\,{{[\nabla_{\rm pol}(p_e/n_e e)\times{\bf B}_{\rm pol}]_z}\over{B^2}} 
\nonumber \\
&=& -\, c\,{{E_{y'} B_{\rm pol}}\over{B^2}} \, - \, 
c\,{{\partial_{y'}(p_e/n_e e)\, B_{\rm pol}}\over{B^2}} \, .
\end{eqnarray}
This is a sum of two drifts: the ${\bf E\times B}$-drift due to 
the poloidal electric field $E_{y'}$ and the diamagnetic drift 
due to the electron pressure gradient. In principle, as long
as the electron pressure is isotropic, these two terms can be 
combined by noticing that in a steady state the poloidal electric 
field is electrostatic, ${\bf E}_{\rm pol}=-\nabla\phi_2(x,y)$,
and defining $\tilde{\phi}_2\equiv \phi_2-p_e/n_e e$. Then,
\beq
v_{\perp,z}^{(e)} =
c\, {{(\partial_{y'}\tilde{\phi}_2)\, B_{\rm pol}} \over{B^2}} \, .
\eeq

However, an important point is that the diamagnetic drift is 
actually irrelevant, as far as the motion of field lines is 
concerned. In the presence of the pressure gradient, the field 
line velocity in fact differs from the electron perpendicular
velocity and is given by just the ${\bf E\times B}$ velocity.
Its $z$-component is 
\beq
v_{B,z} = -\, c\, {{E_{y'}\,B_{\rm pol}}\over{B^2}} \, .
\eeq

\begin{figure} [h]
\centerline
{\psfig{file=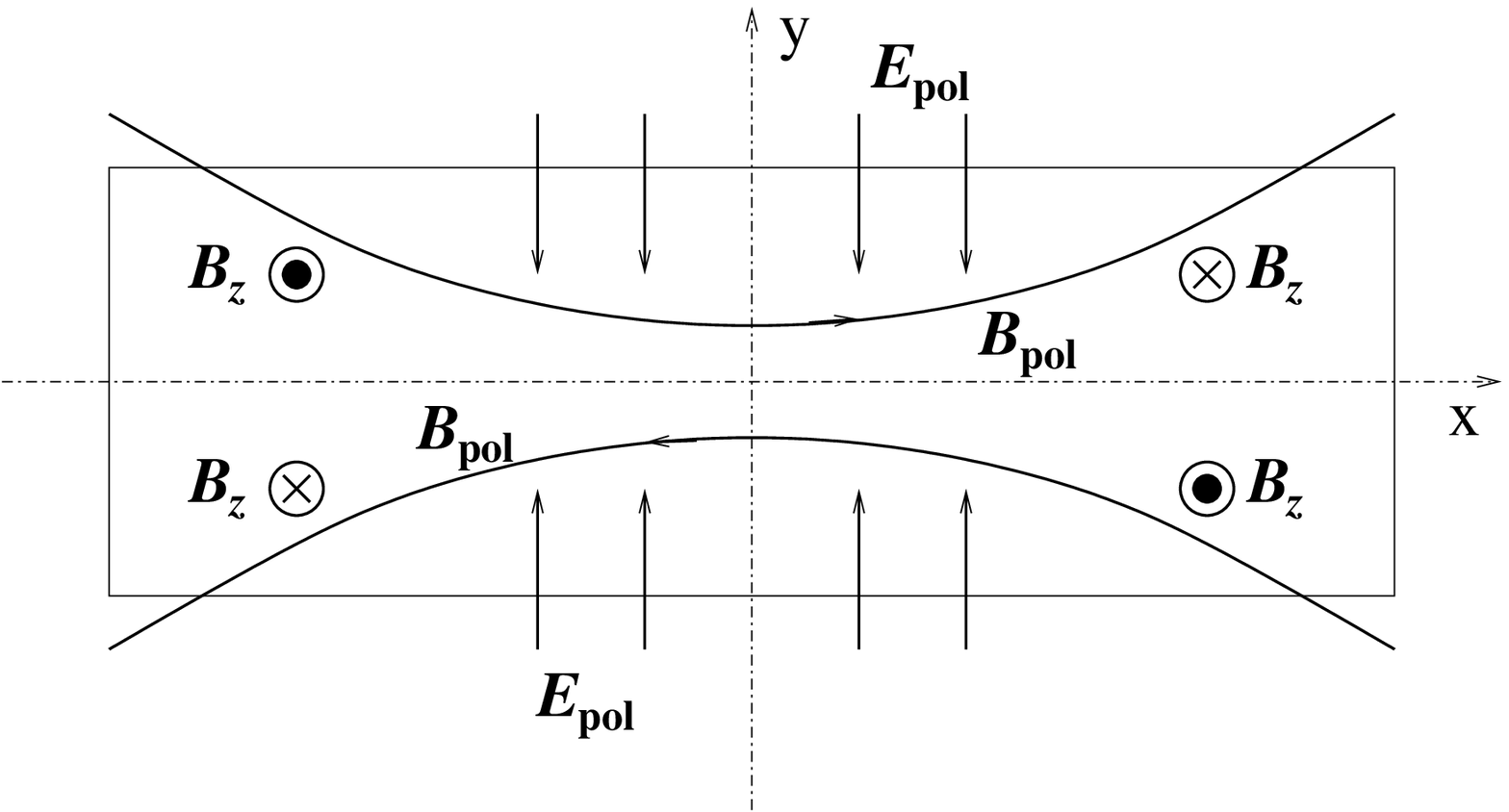,width=5in}}
\caption{Bipolar poloidal electric field.}
\label{fig-Ey}
\end{figure}

Thus, the field lines move toroidally because of~$E_{y'}$ 
that has a bipolar structure (see Fig.~\ref{fig-Ey}).  
The above argument suggests that, instead of saying that electrons pull 
the field lines in the toroidal direction in two-fluid reconnection, it 
is, in a sense, better to say that it is the magnetic field lines that
start moving toroidally and pull the electrons with them. The poloidal 
electric field can therefore be viewed, similarly to the quadrupole 
toroidal magnetic field, as an important signature of Hall reconnection. 
It has in fact been detected with with the {\it Polar spacecraft} in 
the magnetopause~\cite{Mozer-2002}, with the {\it Cluster} spacecraft 
in reconnection regions in the Earth magnetotail~\cite{Wygant-2005,Borg-2005} 
and in the SSX experiment~\cite{Matthaeus-2005}; it has also been seen in 
numerical simulations~\cite{Shay-1998b,Arzner-2001,Pritchett-2001}.
It is this electric field that pulls ions into the reconnection layer; 
as they move across the layer, they pick up the elecrostatic potential
difference of the order of $\delta E_{y'}$. This potential difference 
is large enough to accelerate ions (in the $y$-direction) up to about 
Alfv\'en speed. As a result, the ion $v_y$-distribution at the center 
of the reconnection layer is well represented by two counter-streaming 
beams, which agrees both with numerical particle simulations~\cite
{Shay-1998b,Arzner-2001} and with spacecraft measurements~\cite{Wygant-2005}. 
Effectively, this process can be interpreted as a strong ion heating, 
providing the pressure support for the layer. In addition, ion collisions 
(with particles or with waves) may quickly isotropize the ion distribution 
function, leading to a true ion heating. The quadrupole toroidal magnetic 
field also plays an important role in the poloidal ion motion; in particular, 
as the ions are accelerated into the layer, the Lorentz force due to this 
field bends ion trajectories in the $x$-direction and thus leads to the 
ejection of ions out of the reconnection region.

For reference, we give an expression for the bipolar 
electric field for our simple X-point configuration example.
To derive this expression, we make use of the toroidal 
electron velocity $v_z^{(e)}$, computed in~Sec.~\ref{subsec-v_z}. 

First, from the $y'$-component of the ideal eMHD Ohm's law, 
we can write (neglecting electron pressure)
\beq
c E_{y'} = - v_z^{(e)} B_{x'} + v_{x'}^{(e)} B_z =
- v_z^{(e)} B_{\rm pol} + v_{\rm pol,\parallel} B_z = 
- B_{\rm pol}\, (v_z^{(e)} - \lambda B_z) \, ,
\label{eq-Ey-Xpoint-1}
\eeq
where we express $v_{x'}\equiv v_{\rm pol,\parallel}$ as 
$\lambda B_{\rm pol}$ as it is done in Appendix~\ref{sec-appendix-A}.
In the case of the simple X-point configuration of~Sec.~\ref{subsec-Xpoint},
we have at our disposal explicit expressions for all the ingredients that 
enter the above equation. Thus, $B_{\rm pol}$ is approximately equal 
to~$B_x$ given by equation~(\ref{eq-Xpoint-Bx}); $v_z^{(e)}$ 
is given by equation~(\ref{eq-Xpt-v_z-2}), $\lambda$ by~(\ref
{eq-appA-lambda-3}), and $B_z$ by~(\ref{eq-Bz-2}). Putting it 
all together, we obtain
\beq
E_{y'}(\bar{x},\bar{y}) = 
-\, {Q\over 2}\, |E_z|\, {L\over\delta}\, {1\over{2\bar{\Psi}'}}\, 
\biggl({{\bar{y}}\over{\bar{x}_\delta(\bar{\Psi}')}}\, \log\biggl| 
{{1+\bar{x}_\delta(\bar{\Psi}')}\over{1-\bar{x}_\delta(\bar{\Psi}')}}\biggr|\,
- \bar{x}\, \log\biggl|{{\bar{y}+\bar{x}}\over{\bar{y}-\bar{x}}}\biggr|
\biggr) \, .
\label{eq-Ey-Xpoint-2}
\eeq
We see that, because of the $L/\delta$ factor, this poloidal
field can be considerably larger than the toroidal electric field.


\section{Conclusions}
\label{sec-conclusions}

In this paper we have investigated the structure of a reconnection
layer in the Hall-MHD regime, in which electrons are well-magnetized 
inside the layer, whereas ions are not. Specifically, we have addressed 
the issue of how the quadrupole pattern of an out-of-plane (toroidal) 
magnetic field is generated inside a Hall-MHD reconnection region.
This quadrupole pattern is commonly seen as an important feature of 
two-fluid physics that is at work in the reconnection process whenever 
the resistivity is small. It has been routinely observed both in numerical 
simulations and in space, and has recently been confirmed in a dedicated 
laboratory experiment~\cite{Ren-2005}. In our view, this quadrupole 
pattern arises most naturally via the following mechanism. 

Let us follow a flux tube as it enters the reconnection layer 
from the upstream region. As it moves deeper into the layer, 
the in-the-plane (poloidal) field in the central part of the 
tube weakens and so its cross-sectional area expands. This 
does not affect the ions very much. Let us assume that their 
density is constant throughout the inner part of the reconnection layer. 
Then, owing to charge neutrality, the electron density also has to be 
constant. Therefore, since the central part of the flux tube is expanding, 
electrons have to flow in into the layer along the poloidal magnetic 
field. Similarly, in the region downstream of the X-point, the flux 
tube is leaving the layer and so its cross-sectional area contracts. 
The electrons then are forced to flow along the poloidal field out 
of the layer. We thus obtain a circulating pattern of the electron 
current. In turn, it gives rise, through Ampere's law, to a toroidal 
magnetic field that automatically has a quadrupole structure. A more 
detailed qualitative description of this process is presented 
in~Sec.~\ref{subsec-idea}.

We find that the most elegant and effective way to quantitatively 
analyze the behavior of the system is in terms of the volume-per-flux 
integral~$V(x,\Psi)$, which has a nice property that it is determined 
entirely by the poloidal magnetic field structure, $\Psi(x,y)$. 
We show that both the electron stream function~$\Phi_e$ and 
the toroidal magnetic field~$B_z$ are just proportional to~$V$.
Thus, once the poloidal field structure is specified, the poloidal 
electron velocity and all three components of the magnetic field 
are easily determined just by computing one integral, i.e., without 
solving any partial differential equations.
In particular, we find that, as long as poloidal ion currents are neglected, 
the toroidal magnetic field is constant along electron streamlines.  
This means that, within a pure eMHD framework, the toroidal field 
cannot be produced! Instead, it has to come from the outer regions
of the reconnection layer, where ion currents are not negligible.

We also find that the toroidal magnetic field is highly concentrated near 
the magnetic separatrix. We obtain explicit expressions
for a simple X-point configuration and show that $B_z$ has a logarithmic 
singularity at the separatrix. In reality, of course, the vertical pressure 
balance condition would prevent the toroidal field from being larger that 
the outside poloidal magnetic field~$B_0$. This means that the pressure 
associated with the toroidal magnetic field becomes dynamically important 
near the separatrix and hence the poloidal field structure must be such 
as to keep the toroidal field finite. In addition, the singularity at 
the separatrix is removed by electron inertia. In order to estimate 
the electron inertial term in the toroidal component of the 
generalized Ohm law, however, one needs to know the toroidal electron 
velocity. To determine it, we consider how the full three-dimensional 
shape of a field line changes with time as the field line is advected 
into the layer. From this we deduce the toroidal electron velocity and 
hence estimate how rapidly the electron inertial term grows near the 
separatrix. This enables us to estimate size of the region around the 
separatrix where the electron inertia is not negligible. 

It should be remarked that, in spite of the fact that our calculation
diverges near the separatrix, it is perfectly valid in the upstream 
and downstream regions away from the separatrix. In fact, all our
integrations are carried out from the~$x$ and~$y$ axes towards the 
separatrix and do not cross it.

Finally, we consider the well-known explanation of how the quadrupole 
toroidal field is produced (see, e.g., Ref.~\cite{Mandt-1994}). This 
explanation invokes the differential stretching of poloidal field lines 
by a non-uniform electron flow in the toroidal direction. So a natural 
question to ask is: what makes the electrons move in the toroidal
direction inside the layer? We argue that the toroidal electron
velocity is in fact the sum of the ${\bf E\times B}$ drift, associated 
with the bipolar poloidal electric field that points into the layer, 
and the diamagnetic drift due to electron pressure gradient. However, 
the latter does not lead to any motion of the field lines, and so the 
entire field-line stretching has to be attributed solely to the bipolar 
poloidal electric field. This illustrates the usefulness of this bipolar 
electric field as an important marker for two-fluid effects in the 
reconnection process. We also note that this electric field plays an 
important role in ion dynamics inside the reconnection layer. Namely, 
it is this field that is responsible for accelerating ions towards
the midplane, leading to two counter-streaming ion beams and thus 
to an effective ion heating. The quadrupole toroidal magnetic field 
also plays an important role in ion dynamics as it diverts the two 
beams out of the layer (via the Lorentz force), thereby creating 
the expected stagnation-point pattern for the ion flow.


\acknowledgements
We would like to thank M.~Yamada, H.~Ji, Y.~Ren, A.~Bhattacharjee, 
J.~Drake, and M.~Shay for stimulating and encouraging discussions. 
We are also very grateful to F.~ Mozer and to the anonymous referee 
for a number of very useful comments and suggestions that have 
improved the paper.

This research has been supported by the National Science Foundation 
Grant No.~PHY-0215581 (PFC: Center for Magnetic Self-Organization in 
Laboratory and Astrophysical Plasmas).


\appendix

\section{Alternative derivation of the electron velocity field in 
the simple X-point geometry}
\label{sec-appendix-A}

In this appendix we present an alternative derivation of the poloidal 
electron velocity (and hence the toroidal magnetic field) for our simple 
X-point magnetic structure described by equations~(\ref{eq-Psi-Xpoint}) 
and~(\ref{eq-Psi'}). We make the same two basic assumptions: 
frozen-in law for the poloidal electron flow (which is not altered 
by the electron pressure) and incompressibility. Let us split the 
{\it poloidal} electron velocity field into two parts: parallel and 
perpendicular with respect to the {\it poloidal} magnetic field. 
According to equation~(\ref{eq-v_pol_perp-2}), the perpendicular 
velocity is given by 
\beq
{\bf v}_{\rm pol,\perp} = 
c\, {{E_z\hat{z}\times{\bf B}_{\rm pol}}\over{B_{\rm pol}^2}} \, .
\label{eq-appA-v_perp}
\eeq

In terms of the scaled variables introduced in~Sec.~\ref{subsec-Xpoint},
we can write the~$x$ and~$y$ components of this velocity as 
\begin{eqnarray}
\bar{v}_{\perp, x} &=& {{v_{\perp, x}}\over L} = |\bar{E}|\, 
\biggl({\delta\over L}\biggr)^2 \, {\bar{x}\over{\bar{B}_{\rm pol}^2}} 
\label{eq-appA-v_perp-x}    \, ,       \\
\bar{v}_{\perp, y} &=& {{v_{\perp, y}}\over\delta} = 
-|\bar{E}|\, {\bar{y}\over{\bar{B}_{\rm pol}^2}} \simeq 
-|\bar{E}|\, {1\over{\bar{y}}}
\label{eq-appA-v_perp-y}     \, .
\end{eqnarray}

The parallel part of the poloidal velocity can be written as
\beq
{\bf v}_{\rm pol,\parallel}=\lambda(\bar{x},\bar{y})\,\bar{\bf B}_{\rm pol}\, .
\label{eq-appA-v_par}
\eeq

The total poloidal velocity field has to satisfy the incompressibility
constraint,
\beq
\nabla \cdot {\bf v} = \nabla \cdot {\bf v}_{\rm pol,\perp} + 
\nabla \cdot {\bf v}_{\rm pol,\parallel} = 0 \, .
\label{eq-appA-div-v=0}
\eeq

Using expressions~(\ref{eq-appA-v_perp-x})---(\ref{eq-appA-v_perp-y}), 
we can write the divergence of the perpendicular poloidal velocity, 
to lowest order in~$\delta/L$, as
\beq
\nabla \cdot {\bf v}_{\rm pol,\perp} = 
\bar{\nabla} \cdot \bar{\bf v}_{\rm pol,\perp} \simeq 
\partial_{\bar{y}}\, \bar{v}_{\perp,y} = {|\bar{E}|\over{\bar{y}^2}} \, .
\label{eq-appA-div-v_perp}
\eeq

The divergence of the parallel velocity can be written as
\beq
\nabla \cdot {\bf v}_{\rm pol,\parallel} = 
\nabla\cdot(\lambda{\bf B}_{\rm pol}) = 
{\bf B}_{\rm pol}\cdot\nabla \lambda = 
B_{\rm pol}\, \partial_{l_{\parallel}} \lambda   \, ,
\label{eq-appA-div-v_par}
\eeq
where $l_{\parallel}$ is the path length along a poloidal field line.
Since the divergence of the total velocity must be zero, we get an 
expression for~$\lambda$:
\beq
\lambda (\bar{\Psi},\bar{x}) = 
-\int\limits_0^{l_{\parallel}(\bar{x})}\,(\nabla\cdot{\bf v}_{\rm pol,\perp})
\, {dl_{\parallel}\over{B}_{\rm pol}} \biggl|_{\bar{\Psi}} = 
-\, {L\over{B_0}}\, \int\limits_0^{\bar{x}} \,
(\nabla\cdot{\bf v}_{\rm pol,\perp})\, 
{d\bar{x}\over{\bar{B}_x}} \biggl|_{\bar{\Psi}} \, .
\label{eq-appA-lambda-1}
\eeq

Using our expression~(\ref{eq-appA-div-v_perp}) 
for $\nabla\cdot{\bf v}_{\rm pol,\perp}$, we get
\beq
\lambda (\bar{\Psi},\bar{x}) = 
-\, {L\over{B_0}}\, |\bar{E}| \, \int \limits_0^{\bar{x}}\, 
{d\bar{x}\over{\bar{y}^3(\bar{x},\bar{\Psi})}} \biggl|_{\bar{\Psi}} \, .
\label{eq-appA-lambda-2}
\eeq

Using equation~(\ref{eq-shape-yx}) for the field line shape, 
$\bar{y}(\bar{x},\bar{\Psi})$, we have
\beq
\lambda(\bar{\Psi},\bar{x}) = 
-\, {L\over{B_0}}\, |\bar{E}| \, \int\limits_0^{\bar{x}}\, 
{d\bar{x}\over{(2\bar{\Psi}+\bar{x}^2)^{3/2}}} \biggl|_{\bar{\Psi}} = 
-\, {L\over{B_0}}\, {|\bar{E}|\over{2\bar{\Psi}}}\, {\bar{x}\over{\bar{y}}}\, .
\label{eq-appA-lambda-3}
\eeq

Correspondingly, the components of the parallel velocity are
\begin{eqnarray}
\bar{v}_{\parallel,x} &=& {{v_{\parallel,x}}\over L} =
{\lambda\over L}\, B_x = -\, {|\bar{E}|\over{2\bar{\Psi}}}\, \bar{x}
\label{eq-appA-v_par-x}  \, ,        \\
\bar{v}_{\parallel,y} &=& {{v_{\parallel,y}}\over\delta} =
{\lambda\over\delta}\, B_y = -\, {|\bar{E}|\over{2\bar{\Psi}}}
\, {\bar{x}^2\over\bar{y}} 
\label{eq-appA-v_par-y}      \,   .
\end{eqnarray}

By combining this result with the components of the perpendicular
velocity, we finally get
\begin{eqnarray}
\bar{v}_{x} &=& \bar{v}_{\parallel, x} + \bar{v}_{\perp, x} \simeq
\bar{v}_{\parallel, x} = -\, {|\bar{E}|\over{2\bar{\Psi}}}\, \bar{x} \, ,\\
\bar{v}_{y} &=& \bar{v}_{\parallel, y} + \bar{v}_{\perp, y} = 
-\, {|\bar{E}|\over\bar{y}}\, \biggl({\bar{x}^2\over{2\bar{\Psi}}}+1\biggr)=
-\, {|\bar{E}|\over{2\bar{\Psi}}}\, \bar{y} \, , 
\end{eqnarray}
--- in complete agreement with our calculation in~Sec.~\ref{subsec-Xpoint}!


\section{2D Stationary Ideal Electron MHD: General Formalism}
\label{sec-appendix-B}

In this section we describe a general formalism for analyzing 
a translationally-symmetric electron-MHD system in a steady state. 
We assume here that the electron density is uniform. In addition, 
we neglect any ion currents, both toroidal and poloidal; this  
assumption is valid if, for example, the ion temperature is negligible.
We also neglect electron inertia; however, we do include an isotropic 
electron pressure in our equations.

In full generality, the system is described by three vector fields: 
${\bf B}$, ${\bf v}^{(e)}$, and~${\bf E}$, and a scalar electron 
pressure~$p_e$; it is thus quite complicated. However, these fields 
are not all independent of each other. It turns out that the eMHD 
framework is so constraining that, with the help from the 
time-stationarity and translational symmetry conditions, the
magnetic and electron velocity fields can be expressed in terms of 
only a single one-dimensional function and a constant. In the following, 
we outline how this is done.

First, from $\nabla\cdot{\bf B}=0$ and the translational symmetry 
with respect to~$z$, the magnetic field can be represented by two 
functions: the poloidal flux function $\Psi(x,y)$ (which in this 
section is measured~from the X-point $x=0=y$) and the toroidal 
field~$B_z(x,y)$:
\beq
{\bf B} = B_z \hat{z} + {\bf B}_{\rm pol} = 
B_z \hat{z} + \nabla\times [\Psi \hat{z}] =
B_z \hat{z} + [\nabla\Psi \times \hat{z}]    \, .
\label{eq-AppB-B}
\eeq
The poloidal magnetic field components are
\begin{eqnarray}
B_x &=&  \partial_y \Psi 
\label{eq-AppB-Bx}       \, ;          \\
B_y &=& -\, \partial_x \Psi 
\label{eq-AppB-By}    \, .
\end{eqnarray}

Next, the electron velocity ${\bf v}^{(e)}$ is completely determined 
in terms of the magnetic field by Ampere's law:
\beq
{\bf v}^{(e)}= -\,{j_e\over{n_e e}} = -\, {1\over D}\, \nabla\times{\bf B}\, ,
\label{eq-AppB-v}
\eeq
where $D\equiv 4\pi n_e e/c$. In particular,
\begin{eqnarray}
v_z^{(e)} &=& -\, {1\over D}\, [\nabla\times[\nabla\times(\Psi\hat{z})]]_z =
{1\over D}\, \nabla^2 \Psi 
\label{eq-AppB-vz}         \, , \\
{\bf v}_{\rm pol}^{(e)} &=& -\, {1\over D}\, [\nabla\times(B_z\hat{z})] =
 -\, {1\over D}\, [\nabla B_z\times \hat{z}] 
\label{eq-AppB-vpol}  \, .
\end{eqnarray}
Since the density is uniform, $D$ is constant.

Now let us turn to the electric field~${\bf E}$.
Generally speaking, the overall global magnetic 
configuration evolves as a result of reconnection. 
In particular, there is a continuous transfer of
poloidal magnetic flux through the X-point and
hence there is a non-zero toroidal electric field
at that point. This electric field is a measure 
of the reconnection rate; it is inductive in nature.
However, if the reconnection process is changing 
quasi-statically on the dynamical (Alfv\'en) time 
scale, then locally, inside and around the layer, 
the magnetic field is essentially stationary, 
$\partial_t{\bf B}=0$. Faraday's law then gives 
$\nabla\times{\bf E}=0$. Because of the translational 
symmetry the toroidal electric field then has to be 
uniform; then, the poloidal electric field has to be 
potential:
\begin{eqnarray}
E_z(x,y) &=& {\rm const} \, ; \\
{\bf E}_{\rm pol}(x,y) &=& -\nabla\phi_2(x,y) \, .
\label{eq-def-phi_2}
\end{eqnarray}
Thus, the full three-dimensional electric field is described 
by a constant inductive reconnection field~$E_z$ and a 2D 
electrostatic potential~$\phi_2(x,y)$.

This electric field is tied to the magnetic and velocity fields 
by the generalized Ohm law, i.e., the electron equation of motion. 
Neglecting the inertial and resistive terms but taking into account 
the Hall term and the electron pressure gradient term, this law becomes
\beq
{\bf E} + {1\over c}\, [{\bf v}^{(e)}\times{\bf B}] +
{{\nabla p_e}\over{n_e e}}=0      \, .
\eeq
By taking the vector product of this equation with ${\bf B}$, 
we get an expression for the perpendicular electron velocity:
\beq
{\bf v}_{\perp}^{(e)} = c\, {{{\bf E}\times{\bf B}}\over{B^2}} +
{c\over n_e e}\, {{\nabla p_e \times{\bf B}}\over{B^2}} \, .
\eeq
The first term in this equation is the ${\bf E}\times{\bf B}$
drift and the second term is the diamagnetic drift. Thus, the 
main effect of the pressure gradient term is the diamagnetic current. 
This current is in addition to that associated with the guiding center 
motion due to the ${\bf E}\times{\bf B}$ drift and needs to be included 
in the total current that one substitutes in Ampere's law.

Provided that the electron density is uniform and the electron
pressure tensor is isotropic, the diamagnetic currents do not 
lead to any substantial change in the mathematical structure of 
our formalism and are easily incorporated into our analysis. 
Indeed, combine the electric and pressure terms in Ohm's law 
into one by defining the modified electric field vector:
\beq
\tilde{\bf E} \equiv {\bf E} + {1\over{n_e e}}\, \nabla p_e \, .
\label{eq-def-E-tilde}
\eeq
As we mentioned earlier, the poloidal electric field is potential, 
${\bf E}_{\rm pol}=-\nabla\phi_2$. Then, since we also assume that 
the electron density is uniform in space, $\tilde{\bf E}_{\rm pol}$ 
is also potential, i.e.,
\beq
\tilde{\bf E} = E_z\hat{z} - \nabla\tilde{\phi}_2 \equiv 
E_z\hat{z} - \nabla\biggl(\phi_2 -\, {p_e\over{n_e e}}\biggr) \, ,
\label{eq-E-phi-modified}
\eeq
and so ${\bf v}_{\perp}^{(e)}=c\,[\tilde{\bf E}\times{\bf B}]/B^2$. 

This means that we can take into account the diamagnetic currents 
resulting from the electron pressure gradient simply by working 
with~$\tilde{\phi}_2$ and~$\tilde{\bf E}$ instead of~$\phi_2$ 
and~${\bf E}$. 
Note that one thus cannot really distinguish between the electrostatic 
potential and the electron pressure, and hence between the ${\bf E\times B}$ 
drift and the diamagnetic drift, within the eMHD framework. This degeneracy 
however is not important; in particular, the magnetic and velocity fields 
can still be uniquely determined.

The generalized Ohm law can now be written as
\beq
\tilde{\bf E} = -\, {1\over c}\, [{\bf v}^{(e)}\times{\bf B}] \, .
\label{eq-appB-Ohm}
\eeq
Using equations~(\ref{eq-AppB-vz})--(\ref{eq-AppB-vpol}) for ${\bf v}^{(e)}$ 
and equation~(\ref{eq-AppB-B}) for~${\bf B}$, we can then express the 
components of~$\tilde{\bf E}$ as
\begin{eqnarray}
\tilde{E}_z &=& E_z = 
{1\over{cD}}\, [\nabla B_z \times \nabla\Psi]_z = {\rm const} 
\label{eq-appB-Ez}     \, ,     \\
\tilde{\bf E}_{\rm pol} &=& -\, {1\over{cD}}\, 
\bigl(\nabla^2 \Psi \nabla\Psi + B_z \nabla B_z \bigr) 
\label{eq-appB-Epol} \, .
\end{eqnarray}

But expressing $\tilde{\bf E}_{\rm pol}$ in terms of the 2D 
potential~$\tilde{\phi}_2(x,y)$ by equation~(\ref{eq-E-phi-modified}), 
we see that $(\nabla^2\Psi)\nabla\Psi=\nabla g(x,y)$, where 
$g\equiv cD\tilde{\phi_2}-{B_z^2}/2$.
By taking the curl of this equation, we obtain
$\nabla\times(\nabla^2\Psi \nabla\Psi)=
\nabla(\nabla^2\Psi)\times\nabla\Psi=0$,
and hence ${\bf B}_{\rm pol}\cdot\nabla(\nabla^2\Psi)=0$.
That is, the toroidal current, $j_z \sim \nabla^2\Psi$,
has to be constant along the field lines, and so must be
a function of~$\Psi$ only: 
\beq
\nabla^2\Psi = F(\Psi) \, .
\label{eq-AppB-jz=F(Psi)} 
\eeq
This equation is a necessary condition for the system to have a 
stationary solution, provided that the ion currents are neglected. 
This result is consistent with our earlier (Sec.~\ref{subsec-Bz=Phi=V}) 
finding based on the analysis of the toroidal component of the 
eMHD magnetic induction equation, ${\bf B}_{\rm pol}\cdot\nabla v_z
={\bf v}_{\rm pol}\cdot\nabla B_z$. 
Notice that, in the main part of the paper we considered a general 
poloidal field configuration which could be supported by both ion and 
electron toroidal currents and thus equation~(\ref{eq-AppB-jz=F(Psi)}) 
did not need to be satisfied. However, the poloidal field structure of 
our specific example of Sec.~\ref{subsec-Xpoint} does satisfy this equation, 
which means that it could be produced purely by electron toroidal currents 
with no ion contribution.

Once condition~(\ref{eq-AppB-jz=F(Psi)}) is satisfied, 
we can use the above formalism to compute, one by one, 
all the other electro-magnetic quantities. To do this
for a given function~$F(\Psi)$, one first solves the 
Poisson equation~(\ref{eq-AppB-jz=F(Psi)}) to find~$\Psi(x,y)$, 
and then computes the poloidal magnetic field~${\bf B}_{\rm pol}$ 
and the toroidal velocity~$v_z$ using equations~(\ref{eq-AppB-B})
and~(\ref{eq-AppB-vz}). Next, one uses equation~(\ref{eq-appB-Ez}),
supplemented in the upstream region by the boundary condition 
$B_z(x=0,y)=0$ [and by~$B_z(x,y=0)=0$ in the downstream region],
to calculate the toroidal magnetic field $B_z$. Indeed, the meaning 
of this equation is that the rate of change of the toroidal field 
along a poloidal field line is equal to~$cDE_z/B_{\rm pol}$:
\beq
{\bf B}_{\rm pol} \cdot\nabla B_z = 
[\nabla\Psi\times\hat{z}]\cdot\nabla B_z = 
[\nabla B_z \times \nabla\Psi]\cdot\hat{z} = cDE_z = {\rm const} \, .
\eeq
Therefore, using the symmetry boundary condition at the $y$-axis,
the toroidal field can be immediately obtained by integration along 
the field line:
\beq
B_z(x,\Psi) = cDE_z V(x,\Psi) \equiv 
cDE_z \int\limits_0^x \, {dl_{\rm pol}\over{B_{\rm pol}(l_{\rm pol},\Psi)}}=
cDE_z \int\limits_0^x \, {dx'\over{B_x(x',\Psi)}} \, .
\eeq
Finally, one uses equation~(\ref{eq-AppB-vpol}) to determine 
the poloidal electron velocity and equation~(\ref{eq-appB-Epol})
to determine the modified poloidal electric field.

This concludes the solution of the problem for a given poloidal field. 
The question of what determines the poloidal magnetic field structure 
lies beyond the scope of this paper. Here we just would like to remark 
that this structure is going to be affected by the pressure associated 
with the toroidal magnetic field. Specifically, since the toroidal field 
is strongest near the separatrix, it will push the flux surfaces apart, 
resulting in a weaker poloidal field near the separatrix. This effect
has to be taken into account self-consistently.



\newpage

\newpage

\newpage

\begin{thebibliography}{99}

\bibitem{Yamada-1997}
M.~Yamada, H.~Ji, S.~Hsu, T.~Carter, R.~Kulsrud, N.~Bertz, F.~Jobes, 
Y.~Ono, and F.~Perkins, Phys. Plasmas, {\bf 4}, 1936 (1997).

\bibitem{Sweet-1958} 
P.~A.~Sweet, in {\it Electromagnetic Phenomena in Cosmical Physics}, 
ed. B.~Lehnert, (Cambridge University Press, New York, 1958), p.~123.

\bibitem{Parker-1957} 
E.~N.~Parker, J. Geophys. Res., {\bf 62}, 509 (1957).

\bibitem{Parker-1963} 
E.~N.~Parker, ApJ Supplement, {\bf 8}, 177 (1963).

\bibitem{Petschek-1964} 
H.~E.~Petschek, in {\it AAS-NASA Symposium on Solar Flares}, 
(National Aeronautics and Space Administration, Washington, DC, 1964), 
NASA SP50, 425.

\bibitem{Biskamp-1986} 
D.~Biskamp, Phys. Fluids {\bf 29}, 1520 (1986).

\bibitem{Scholer-1989} 
M.~Scholer, J. Geophys. Res., {\bf 94}, 8805 (1989)

\bibitem{Uzdensky-2000} 
D.~A.~Uzdensky and R.~M.~Kulsrud, Phys. Plasmas, {\bf 7}, 4018 (2000).

\bibitem{Kulsrud-2001} 
R.~M.~Kulsrud, Earth, Planets and Space, {\bf 53}, 417 (2001).

\bibitem{Malyshkin-2005} 
L.~M.~Malyshkin, T.~Linde, and R.~M.~Kulsrud, 
Phys. Plasmas, {\bf 12}, 102902 (2005).

\bibitem{Ugai-1977} 
M.~Ugai and T.~Tsuda, J. Plasma Phys., {\bf 17}, 337 (1977). 

\bibitem{Biskamp-2001} 
D.~Biskamp and E.~Schwarz, Phys. Plasmas, {\bf 8}, 4729 (2001).

\bibitem{Sonnerup-1979} 
B.~U.~\"{O}.~Sonnerup, Magnetic Field Reconnection,
in {\it Solar System Plasma Physics}, vol.~3, ed. L.~T.~Lanzerotti, 
C.~F.~Kennel, and E.~N.~Parker, (North-Holland, New York, 1979) p.~45.

\bibitem{Mandt-1994} 
M.~E.~Mandt, R.~E.~Denton, and J.~F.~Drake, 
Geophys. Res. Lett., {\bf 21}, 73 (1994).

\bibitem{Biskamp-1997} 
D.~Biskamp, E.~Schwarz, and J.~F.~Drake, Phys. Plasmas, {\bf 4}, 1002 (1997).

\bibitem{Shay-1998a} 
M.~A.~Shay and J.~F.~Drake, Geophys. Res. Lett., {\bf 25}, 3759 (1998).

\bibitem{Shay-1998b} 
M.~A.~Shay, J.~F.~Drake, R.~E.~Denton, and D.~Biskamp, 
J. Geophys. Res., {\bf 103}, 9165 (1998).

\bibitem{Terasawa-1983} 
T.~Terasawa, Geophys. Res. Lett., {\bf 10}, 475 (1983).

\bibitem{Wang-2000} 
X.~Wang, A.~Bhattacharjee, and Z.~W.~Ma, J. Geophys. Res., 
{\bf 105}, 27633 (2000). 

\bibitem{Arzner-2001} 
K.~Arzner and M.~Scholer, J. Geophys. Res., {\bf 106}, p.~3827 (2001).

\bibitem{Pritchett-2001} 
P.~L.~Pritchett, J. Geophys. Res., {\bf 106}, p.~25961 (2001).

\bibitem{Breslau-2003} 
J.~A.~Breslau and S.~C.~Jardin, Phys. Plasmas, {\bf 10}, 1291 (2003).

\bibitem{Mozer-2002} 
F.~S.~Mozer, S.~D.~Bale, and T.~D.Phan, 
Phys. Rev. Lett., {\bf 89}, 015002 (2002).

\bibitem{Wygant-2005} 
J.~R.~Wygant, C.~A.~Cattell, R.~Lysak, Y.~Song, J.~Dombeck, 
J.~McFadden, F.~S.~Mozer, C.~W.~Carlson, G.~Parks, E.~A.~Lucek, 
A.~Balogh, M.~Andre, H.~Reme, M.~Hesse, and C.~Mouikis, 
J. Geophys. Res., {\bf 110}, A09206 (2005).

\bibitem{Borg-2005} 
A.~L.~Borg, M.~Oieroset, T.~D.~Phan, F.~S.~Mozer, A.~Pedersen, 
C.~Mouikis, J.~P.~McFadden, C.~Twitty,  A.~Balogh, and H.~Reme, 
Geophys. Res. Lett., {\bf 32}, L19105 (2005).

\bibitem{Ren-2005} 
Y.~Ren, M.~Yamada, S.~Gerhardt, H.~Ji, R.~Kulsrud, and A.~Kuritsin, 
Phys. Rev. Lett., {\bf 95}, 055003 (2005).

\bibitem{Matthaeus-2005} 
W.~H.~Matthaeus, C.~D.~Cothran, M.~Landerman, and M.~R.~Brown, 
Geophys. Res. Lett., {\bf 32}, L23104 (2005).


\end{thebibliography}
\end{document}